\documentclass{aa}
\usepackage{txfonts}
\usepackage{graphicx}
\usepackage{natbib}
\bibpunct{(}{)}{;}{a}{}{,}

\begin{document}
\title{Radio properties of the low surface brightness SNR G65.2+5.7
\thanks{Figures 1 and 2 are available in FITS format at the CDS via
anonymous ftp to cdsarc.u-strasbg.fr (130.79.128.5) or via
http://cdsweb.u-strasbg.fr/cgi-bin/qcat?/A+A/xxx/yyy}}
\author{L. Xiao\inst{1}
        \and W. Reich\inst{2}
    \and E. F\"urst\inst{2}
    \and J. L. Han\inst{1}}


\institute{National Astronomical
Observatories, Chinese Academy of
            Sciences, Jia-20, Datun Road, Chaoyang District, Beijing 100012, China\\
            \email{xl,hjl@bao.ac.cn}
            \and Max-Planck-Institut f\"{u}r Radioastronomie,
             Auf dem H\"ugel 69, D-53121 Bonn, Germany\\
            \email{efuerst,wreich@mpifr-bonn.mpg.de}}

\date{Received / Accepted}

\abstract
{
SNR G65.2+5.7 is one of few supernova remnants (SNRs) that have been optically detected. 
It is exceptionally bright in X-rays and the optical [O \MakeUppercase{\romannumeral 3}]-line.
Its low surface brightness and large diameter ensure that radio observations of SNR G65.2+5.7 are
technically difficult and thus have hardly been completed.
}
{Many physical properties
of this SNR, such as spectrum and polarization, can only be investigated by radio observations.
}
{The $\lambda$11\ cm and $\lambda$6\ cm continuum and polarization observations
of SNR G65.2+5.7 were completed with the Effelsberg 100-m and the Urumqi 25-m
telescopes, respectively, to investigate the integrated spectrum, the spectral index 
distribution, and the magnetic field properties. Archival $\lambda$21\ cm data 
of the Effelsberg 100-m telescope were also used.}
{The integrated flux densities of G65.2+5.7 at $\lambda 11$\ cm and
$\lambda 6$\ cm are $21.9\pm3.1$~Jy and 16.8$\pm$1.8~Jy,
respectively. The power-law spectrum ($S\sim\nu^{\alpha}$) is well fitted by
$\alpha = -0.58\pm0.07$  from 83~MHz to 4.8~GHz.
Spatial spectral variations are small.
Along the northern shell, strong depolarizion is observed at both wavelengths.
The southern filamentary shell of SNR G65.2+5.7 is polarized by as much as 54\% at $\lambda 6$\ cm. 
There is significant depolarization at $\lambda 11$\ cm and confusion with diffuse polarized 
Galactic emission. Using equipartition principle, we estimated the magnetic field strength for the 
southern filamentary shell to be between 20~$\mu$G (filling factor 1) and 50~$\mu$G (filling factor 0.1). 
A faint HI shell may be associated with the SNR.}
{Despite its unusually strong X-ray and optical emission and its very low surface brightness, 
the radio properties of SNR G65.2+5.7 are found to be typical of evolved 
shell-type SNRs. SNR G65.2+5.7 may be expanding in a pre-blown cavity as indicated by a deficit of 
HI gas and a possible HI-shell. 
}

\keywords{supernova remnants -- synchrotron emission -- linear polarization}

\titlerunning{Radio properties of the SNR G65.2+5.7}

\maketitle

\section{Introduction}

SNRs of low radio surface brightness and large diameter are difficult to identify in radio 
continuum surveys. The weakest source detected in our Galaxy is SNR G156.2+5.7 \citep{rfa92}, 
which has a surface brightness $\rm \Sigma_{1GHz}\sim5.8\times 10^{-23} W m^{-2} Hz^{-1} sr^{-1}$,  
and SNR G65.2+5.7 is a similar weak object, whose low surface brightness
may be caused by a low ambient magnetic field strength or a low electron density.
In most cases, the supernova has exploded in a low-density medium, for instance in an interarm 
region or in a pre-existing cavity. In the case of a pre-existing cavity substantial radio emission 
is visible for only a relatively short time interval as the shock collides with the dense shell.

\citet{gkp77} first identified G65.2+5.7 as a large diameter SNR located in the Cygnus region
in an optical emission line survey, where it appeared 
to be exceptionally bright in [O \MakeUppercase{\romannumeral 3}]. This is indicative of high shock
velocities typically for an expanding SNR shell. \citet{rbs79} used
the Effelsberg 100-m telescope at $\lambda 21$cm to map G65.2+5.7
and confirmed its identification as a SNR by its non-thermal emission.

G65.2+5.7 is a bright soft X-ray source for which \citet{skp04}
analysed ROSAT PSPC data, classifying the SNR as a ``thermal composite'' object with a cool, 
dense shell and no X-ray emission. Bright centrally peaked thermal
X-ray emission dominates the interior of these SNRs, although they
are rare. W44 is another example \citep{c99}, which is younger than G65.2+5.7.

Distance estimates were made by various authors. \citet{bml04} used proper
motion and expansion measurements of the remnant's optical filament
edges and obtained 770$\pm$200~pc. In the following, we assume a
distance of 800~pc. The apparent size of G65.2+5.7 is about $4\degr
\times 3\fdg3$ \citep{gkp77}, corresponding to a physical size of 56~pc
$\times$ 46~pc. The centre of the SNR is located about 80~pc above
the Galactic plane. G65.2+5.7 appears to be an evolved object,
although its age is uncertain: It may be as old as $3\times10^{5}$ years
if one assumes an average shock velocity of $\sim$
50~km~s$^{-1}$ \citep{gkp77, rbs79}. Optical observations show the
velocities in the range between 90~km~s$^{-1}$ and 140~km~s$^{-1}$
\citep{mbp02}. ROSAT X-ray data infer shock velocities of about 
400~km~s$^{-1}$ \citep{schau02}, which implies an age of $2.8\times10^{4}$ years.

The Princeton-Arecibo pulsar survey found a pulsar, PSR J1931+30,
in the direction of G65.2+5.7 \citep{cam96}. From its
dispersion measure of 56$\pm11$~cm$^{-3}$pc, a distance of about
3~kpc is inferred, which excludes a physical association with G65.2+5.7.

The only radio observations of G65.2+5.7 have been those by
\citet{rbs79} at $\lambda 21$\ cm and the low-resolution, low-frequency observations by
\citet{kpu94}, reflecting the difficulty in mapping these faint large-diameter 
objects. With the availability of new sensitive and highly stable 
receivers, it is now possible to map this object also at higher
frequencies with arc minute angular resolution and investigate
its spectral properties and magnetic field structure. We describe new
sensitive radio measurements in total power and linear polarization
at $\lambda$11\ cm and $\lambda$6\ cm of G65.2+5.7 and present results
 on the spectral and polarization analysis in Sect.~2. 
A brief discussion of the radio results with respect to observations at other
wavelengths is given in Sect.~3, followed by a summary in Sect.~4.

\section{Observations and result analyses}

Continuum and linear polarization observations of G65.2+5.7 at $\lambda$6\ cm (4800~MHz)
were completed with the 25-m telescope at Nanshan station of the Urumqi Observatory.
The $\lambda$6\ cm receiver is a copy of a receiver used at the Effelsberg 100-m
telescope and was installed in 2004 for the $\lambda$6\ cm polarization 
survey of the Galactic plane \citep{shr07}.
\citet{srh06} described the performance of the receiving
system in some detail. Between 2006 and 2007, nine observations of
G65.2+5.7 were completed, which were centered at $\rm \alpha_{2000}=19^{h}33^{m}6^{s}$,
$\delta_{2000}=31\degr17\arcmin31\arcsec$. The map size, $\Delta$RA $\times$ $\Delta$Dec, was $5\degr
\times 4\fdg5$. The maps were scanned either along the RA- or
Dec-direction with a telescope scan velocity of $4\arcmin$/sec.

\begin{table}
   \caption{Observational parameters}
   \label{obspara}
  {\begin{tabular}{lll}\hline\hline
  Wavelength                      & $\lambda$6\ cm & $\lambda$11\ cm   \\
  Frequency                       & 4800\ MHz    & 2639\ MHz           \\
  Bandwidth                       & 600\ MHz     & 80\ MHz            \\
  HPBW [\arcmin ]                 & 9.5          & 4.4                \\
  apperture efficiency[\%]        & 62           &  53                \\
  beam efficiency[\%]             & 67           &  58                \\
  T$_{\rm sys}$[K]                & 22           &  17                \\
  T$_{B}[K]$/S[Jy]                & 0.164        & 2.52               \\
  Main Calibrator                 & 3C286        & 3C286              \\
  \ \ \   Flux Density            & 7.5\ Jy      & 11.5\ Jy           \\
  \ \ \   Polarization Percentage & 11.3\%       & 9.9\%              \\
  \ \ \   Polarization Angle      & 33$\degr$    & 33$\degr$          \\
  No. of coverages                &  9           & 7                   \\
  pixel integration time [sec]    & 7            & 3.5                  \\
  r.m.s in TP [mK]                & 0.7          &  3.5               \\
  r.m.s in PI [mK]                & 0.3          &  1.7               \\\hline
  \end{tabular}}
\end{table}

The observations of G65.2+5.7 at $\lambda$11\ cm (2639~MHz) were completed with a
receiver installed at the secondary focus of the Effelsberg 100-m telescope in 2005.
They were carried out in the same way as the observations with the Urumqi telescope.
G65.2+5.7 was observed seven times in autumn 2007 mostly in clear sky.
In Table~1, we list the parameters of the $\lambda$6\ cm and
$\lambda$11\ cm observations with the data of the main calibrator 3C286.
 The $\lambda$11\ cm observations were carried out with a
8-channel narrow band polarimeter. Each channel has  a bandwidth of
10~MHz. The centre frequencies are separated by 10~MHz and range from
2604~MHz to 2674~MHz. The 9th broadband-channel has a
bandwidth of 80~MHz and is centered on 2639~MHz. This channel was used 
for all observations in which no narrow band interference was present,
which was the case for the observations of G65.2+5.7.

The data processing was completed in exactly the same way as described by
\citet{xfrh08} for the observations of the SNR S147, an
object of similar size and faintness as G65.2+5.7. In
brief, the individual maps for Stokes I, U, and Q were checked and low quality 
data caused by interference were removed. The baselines for Stokes I
were corrected by subtracting a second order polynomial fit to the
lower emission envelope; this step removed ground emission variations and
also the smooth increase in Galactic emission towards the Galactic
plane evident in the long scans. We
suppressed scanning effects for U/Q maps by applying the ``unsharp masking'' method
described by \citet{sr79}. Finally, all maps were combined using the
PLAIT-algorithm \citep{eg88}.


\subsection{Total intensity and linear polarization maps}

\begin{figure}[!htbp]
\includegraphics[width=0.4\textwidth, angle=-90]{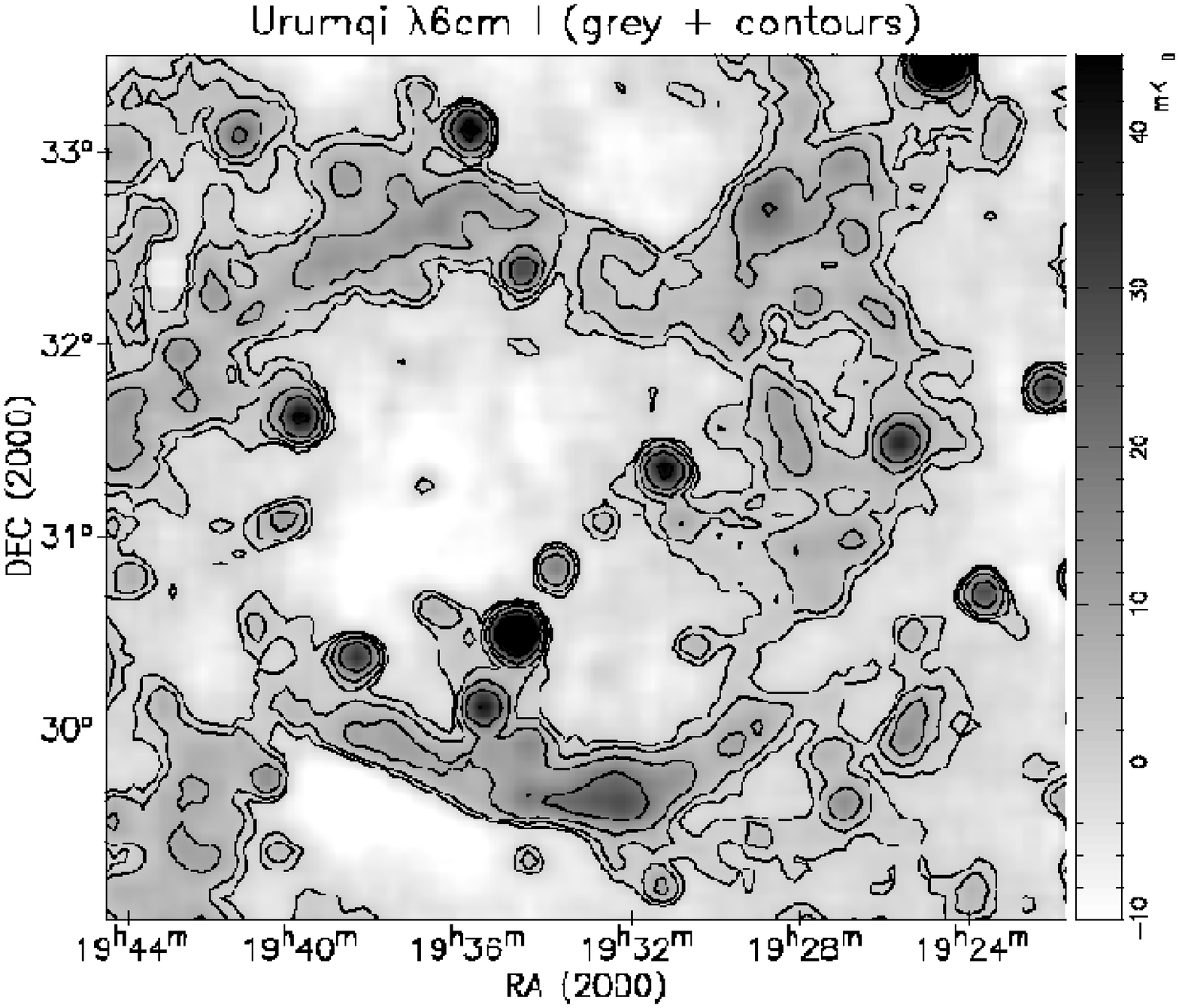}
\includegraphics[width=0.4\textwidth, angle=-90]{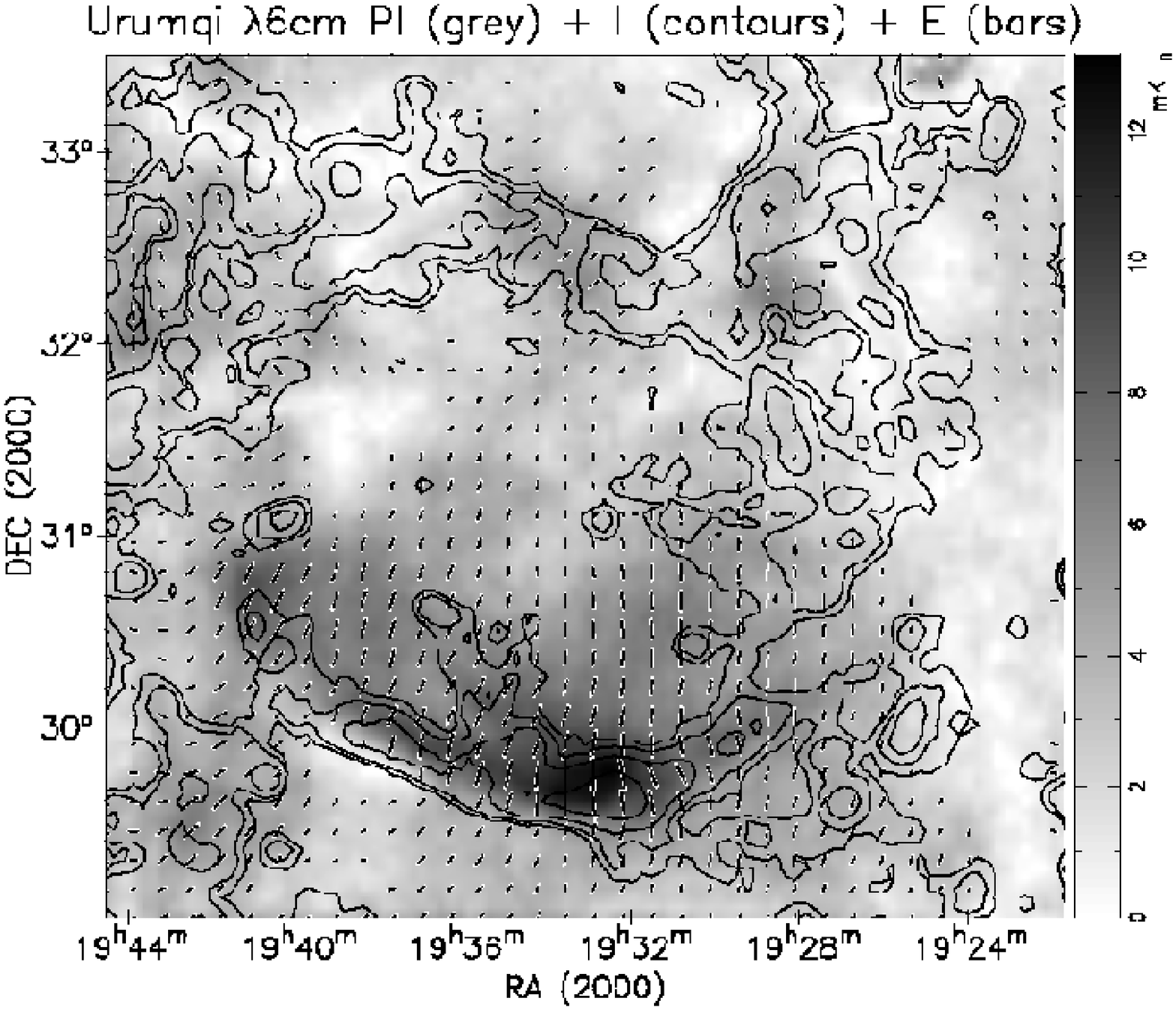}
    \caption{The Urumqi $\lambda$6\ cm map of G65.2+5.7 with an angular resolution of $9\farcm5$.
             The upper panel shows the distribution of total intensities. The lower panel shows 
             polarized intensities
             (PI) in grey superimposed on total intensity contours (with point-like sources subtracted) 
             and bars for E-field direction.
             The length of the bars is proportional to PI. 
             Polarized intensity of 1~mK~T$_{\mathrm B}$ corresponds to a bar-length of $0\farcm86$.
             Both panels show total intensity contours starting at 0~mK~T$_{\mathrm B}$ and increasing 
             $\rm 2^n\times 3\sigma_{I}$, where n=0,1,2,3$\cdots$ and 
             $\rm \sigma_{I}= 0.7$~mK~T$_{\mathrm B}$.}
    \label{6cm}
\end{figure}

\begin{figure}[!htbp]
\includegraphics[width=0.4\textwidth, angle=-90]{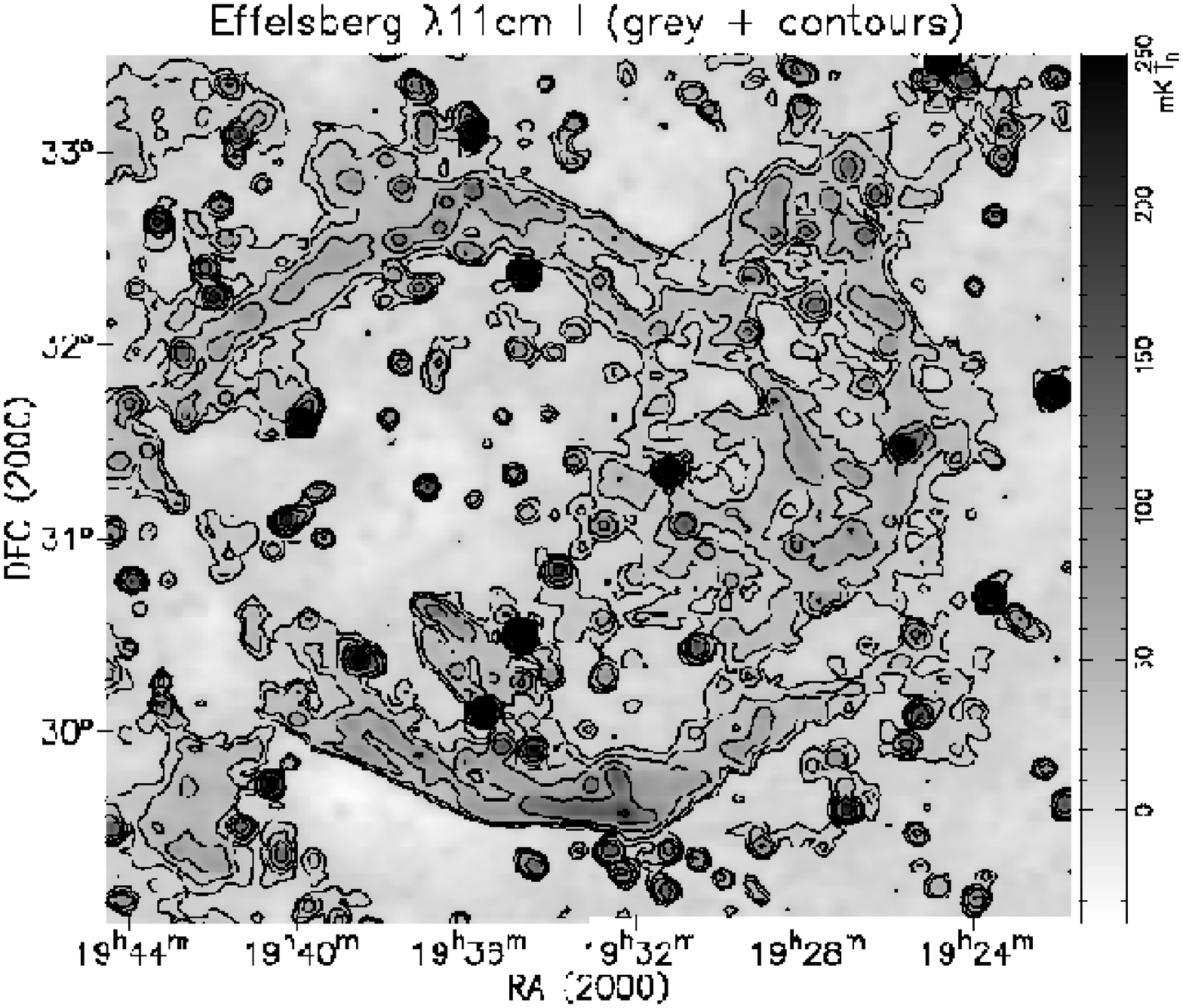}
\includegraphics[width=0.4\textwidth, angle=-90]{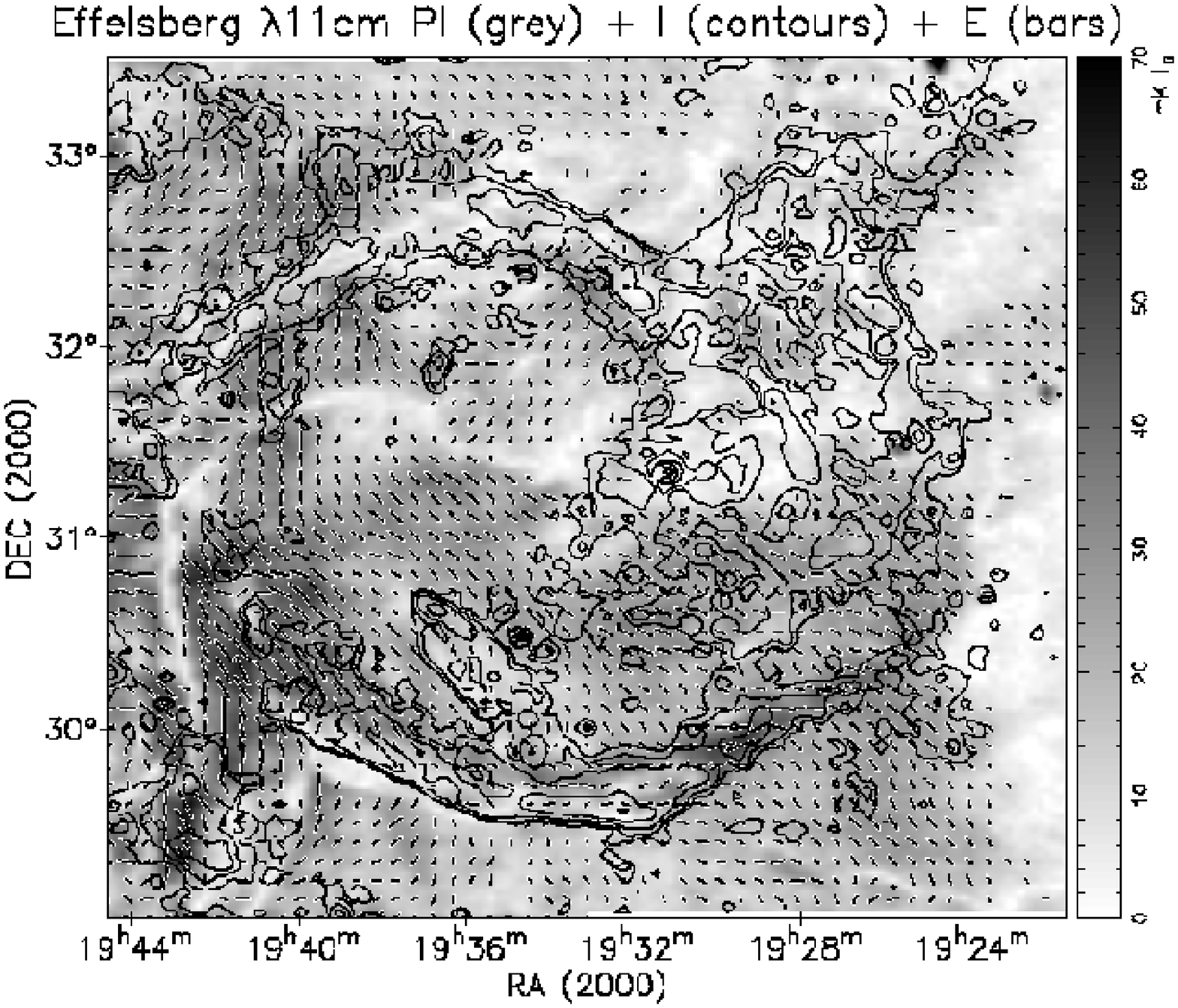}
    \caption{Same as Fig.~1, but for the Effelsberg $\lambda$11\ cm map with an angular resolution 
             of $4\farcm4$.
             Polarized intensity of 1~mK~T$_{\rm B}$ corresponds to a bar-length of $0\farcm25$.
             Total intensity contours start at 21~mK~T$_{\mathrm B}$ and increase in steps of 
             $\rm 2^n\times 3\sigma_{I}$ 
             with n=0,1,2,3$\cdots$, where $\rm \sigma_{I}=3.5$~mk~T$_{\rm B}$.}
    \label{11cm}
\end{figure}

The Urumqi $\lambda$6\ cm total intensity map of G65.2+5.7 and the corresponding 
polarization intensity map superimposed with vectors in the E-field direction 
are shown in Fig.~\ref{6cm}. From low-intensity areas in the maps,
we measured an r.m.s.-noise in total intensity and in linear polarized intensity
of 0.7~mK~T$_{\rm B}$ and 0.3~mK~T$_{\rm B}$, respectively. The $\lambda$11\ cm 
maps are shown in
Fig.~\ref{11cm}. The r.m.s.-noise was measured to be 3.5~mK~T$_{\rm B}$ in total intensity
and 1.7~mK~T$_{\rm B}$ in linear polarized intensity. Because of the higher
angular resolution of the $\lambda$11\ cm map of $4\farcm4$, the filamentary elliptical shell 
is more clearly detected.

\begin{figure}[!htbp]
\includegraphics[width=0.42\textwidth, angle=-90]{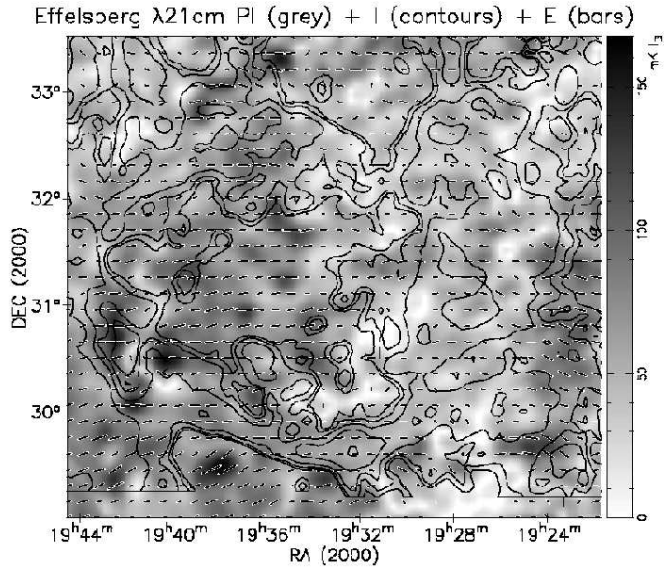}
    \caption{Effelsberg $\lambda$21\ cm polarized intensity map of G65.2+5.7 with an angular resolution 
             of $9\farcm4$. The length of the bars is proportional to PI. 
             Polarized intensity of 1~mK~T$_{\rm B}$ corresponds to a bar-length of $0\farcm075$.
            Contours show total intensities with compact sources removed
            starting at 66~mK~T$_{\rm B}$ and increasing in steps of 
            $\rm 2^n\times 36~mK~T_{\rm B}$ with n=0,1,2,3$\cdots$. 
             }
    \label{21cmpi}
\end{figure}

For the spectral index analysis we also used an archival $\lambda$21\ cm map obtained
with the Effelsberg 100-m telescope in November 1987 at 1408~MHz with a bandwidth
of 20~MHz and an angular resolution of $9\farcm4$. 
These observations were made in due course with a $\lambda$21\ cm Galactic plane
survey in total intensity \citep{rei90}, which covers the plane within an interval of
$\pm 4\degr$ in latitude, so that a small part of G65.2+5.7 is included there.
Polarization data were taken from a section of the ``$\lambda$21\ cm
Effelsberg Medium Latitude Survey'' \citep{rfr04}, which was combined with data from 
the DRAO 1.4~GHz polarization survey~\citep{woll06} to help ensure detection
of any missing large-scale polarized emission. The $\lambda$21\ cm map is shown 
in Fig.~\ref{21cmpi}.

\subsection{The integrated radio spectrum}

To estimate the integrated flux density of the remnant, we
subtracted 20 bright point-like sources by Gaussian fitting within the area of the SNR, 
and a few just outside, 
from the $\lambda$11\ cm map and 16~sources from the $\lambda$6\ cm map.
 The sources are listed in Table~2. The position accuracy of
the two maps is about $7\arcsec$ on average, which was estimated by comparing the source 
positions with those from the NVSS \citep{con98}.

\begin{table*}
\caption{Bright sources in the area of SNR G65.2+5.7 fitted from the 
Effelsberg $\lambda$11\ cm and  Urumqi at $\lambda$6\ cm maps, including
source positions, flux densities from the NVSS at $\lambda$21\ cm, as well as spectral indices. Sources
'outside' G65.2+5.7 are listed for completeness.}
\label{sources}
\begin{tabular}{r c c r@{.}l@{}r@{.}l r@{}l@{}r r@{}l@{}r c c}
\hline\hline
   &NVSS &NVSS &\multicolumn{4}{c}{$S_{NVSS}$} &\multicolumn{3}{c}{$S_{11cm}$} &\multicolumn{3}{c}{$S_{6cm}$} & $\alpha$  &comment \\
   &$\alpha_{2000}$ &$\delta_{2000}$ &\multicolumn{4}{c}{(mJy)} &\multicolumn{3}{c}{(mJy)} &\multicolumn{3}{c}{(mJy)}  &  & \\
\hline
 1 &19 23 25.26 &30 42 35.4 &574 & 4$\pm$&19 & 1 &300 &$\pm$&14 &159 &$\pm$&16 &-0.83$\pm$0.05 & \\
 2 &19 25 13.64 &30 31 18.9 & 70 & 3$\pm$& 2 & 1 & 43 &$\pm$&7  & 31 &$\pm$&5  &-0.81$\pm$0.19 &\\
 3 &19 25 27.29 &31 29 23.9 &384 & 3$\pm$&12 & 6 &222 &$\pm$&12 &171 &$\pm$&9  &-0.64$\pm$0.07 &\\
 4 &19 26 16.04 &32 35 51.6 &105 & 4$\pm$& 3 & 9 & 47 &$\pm$&14 &    &     &   &-1.28$\pm$0.49 &\\
 5 &19 26 40.65 &32 56 55.9 & 67 & 1$\pm$& 3 & 0 & 48 &$\pm$&12 &    &     &   &-0.53$\pm$0.42 &3 unresolved sources\\
 6 &19 26 56.04 &29 37 19.6 &193 & 2$\pm$& 6 & 0 &111 &$\pm$&6  & 67 &$\pm$&17 &-0.84$\pm$0.14 &outside\\
 7 &19 27 11.53 &29 52 54.5 & 51 & 6$\pm$& 2 & 2 & 28 &$\pm$&9  & 24 &$\pm$&5  &-0.77$\pm$0.21 &outside\\
 8 &19 27 33.05 &32 14 01.3 & 79 & 2$\pm$& 2 & 8 & 52 &$\pm$&9  & 43 &$\pm$&8  &-0.72$\pm$0.18 &\\
 9 &19 30 28.65 &30 27 42.0 &113 & 1$\pm$& 3 & 4 & 71 &$\pm$&4  & 37 &$\pm$&4  &-1.03$\pm$0.15 &\\
10 &19 30 46.84 &31 05 59.1 &126 & 2$\pm$& 4 & 4 & 62 &$\pm$&8  & 43 &$\pm$&9  &-0.80$\pm$0.15 &\\
11 &19 31 08.60 &31 22 33.5 &428 & 6$\pm$&12 & 9 &311 &$\pm$&11 &232 &$\pm$&14 &-0.63$\pm$0.05 &11/6cm fit\\
12 &19 31 14.43 &29 12 35.2 & 67 & 7$\pm$& 2 & 1 & 59 &$\pm$&16 & 55 &$\pm$&5  &-0.66$\pm$0.19 &outside\\
13 &19 32 45.08 &31 06 10.7 & 35 & 8$\pm$& 1 & 1 & 38 &$\pm$&8  &    &     &   & 0.09$\pm$0.35 &   \\        
14 &19 33 48.95 &30 51 45.1 & 64 & 7$\pm$& 2 & 0 & 90 &$\pm$&15 & 98 &$\pm$&8  & 0.14$\pm$0.07 &11/6cm fit\\
15 &19 34 19.84 &29 20 07.4 & 65 & 5$\pm$& 2 & 3 & 59 &$\pm$&9  & 37 &$\pm$&8  &-0.84$\pm$0.20 &outside\\
16 &19 34 38.47 &32 24 47.8 &625 & 7$\pm$&18 & 8 &299 &$\pm$&13 &146 &$\pm$&13 &-0.80$\pm$0.04 &\\
17 &19 34 45.20 &30 30 58.8 &235 & 2$\pm$& 7 & 1 &469 &$\pm$&14 &579 &$\pm$&23 &0.35$\pm$0.11 &\\
18 &19 35 34.24 &30 08 00.9 &688 & 7$\pm$&23 & 8 &359 &$\pm$&43 &183 &$\pm$&22 &-0.81$\pm$0.04 &\\
19 &19 35 56.39 &33 08 22.0 & 77 & 0$\pm$& 2 & 3 &216 &$\pm$&10 &250 &$\pm$&16 & 0.24$\pm$0.18 &outside, 11/6cm fit\\
20 &19 36 58.60 &31 17 25.8 &149 & 7$\pm$& 4 & 5 & 93 &$\pm$&9  & 55 &$\pm$&14 &-0.89$\pm$0.16 &\\
21 &19 38 38.31 &30 23 55.4 & 92 & 6$\pm$& 2 & 8 &172 &$\pm$&15 &171 &$\pm$&18 &-0.01$\pm$0.29 &11/6cm fit\\
22 &19 40 03.36 &31 37 40.6 &538 & 5$\pm$&18 & 4 &353 &$\pm$&15 &220 &$\pm$&17 &-0.77$\pm$0.05 &\\
23 &19 40 26.13 &31 06 23.9 &199 & 4$\pm$& 6 & 0 &112 &$\pm$&15 & 61 &$\pm$&8  &-0.94$\pm$0.14 &\\
24 &19 40 39.90 &29 44 46.8 &300 & 1$\pm$&10 & 6 &147 &$\pm$&9  & 73 &$\pm$&17 &-0.92$\pm$0.11 &outside\\
25 &19 42 14.83 &32 15 38.7 &177 & 2$\pm$& 5 & 3 & 98 &$\pm$&12 &    &     &   &-0.93$\pm$0.23 &\\
26 &19 42 29.66 &32 24 46.9 & 83 & 3$\pm$& 2 & 5 & 58 &$\pm$&12 &    &     &   &-0.58$\pm$0.35 &outside\\ 
27 &19 43 00.76 &31 58 14.2 & 76 & 3$\pm$& 2 & 3 & 60 &$\pm$&10 & 43 &$\pm$&16 &-0.79$\pm$0.20 &\\
28 &19 44 05.01 &30 46 45.8 &156 & 6$\pm$& 4 & 7 & 82 &$\pm$&4  & 37 &$\pm$&12 &-0.96$\pm$0.16 &outside\\
\hline
\end{tabular}\\
\end{table*}

We used the method of temperature-versus-temperature (TT) plot \citep{tpk62}
to adjust the base-levels of the entire SNR area at two wavelengths. Both maps
were convolved to a common angular resolution of $10\arcmin$.
In agreement with the TT-plot results for three different shell regions (see Sect.~3.3), 
we found a constant offset of $\rm -6~mK~T_{\rm B}$ at $\lambda$6\ cm. 
At $\lambda$11\ cm, we subtracted a ``twisted'' surface, which was defined by specific
correction values at the four corners (3, 3, $-6$, and 3 mK~T$_{\rm B}$ from lower left, 
right to upper left, right) of the map. 
The integrated flux densities were obtained by integrating the emission enclosed within polygons 
just outside the periphery of the SNR. From variations outside the SNR, we
estimated a remaining uncertainty in the base-level setting of $\rm 1~mK~T_{\mathrm B}$
at $\lambda$6\ cm and $\rm 2\ mK\ T_{\mathrm B}$ at $\lambda$11\ cm.
By assuming an estimated 5\% calibration uncertainty, we obtained
integrated flux densities of 16.8$\pm$1.8\ Jy at
$\lambda$6\ cm and 21.9$\pm$3.1\ Jy at $\lambda$11\ cm.

\begin{figure}[!htbp]
\includegraphics[width=0.32\textwidth, angle=-90]{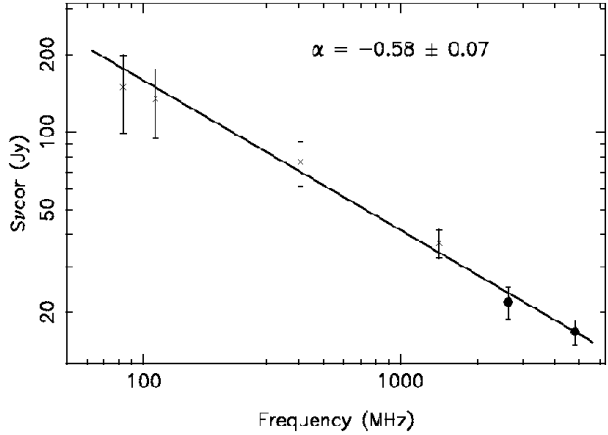}
    \caption{Spectrum of the integrated radio flux densities of G65.2+5.7.
             The spectral index is $\alpha = -0.58\pm0.07$.
             All flux densities were corrected for the contribution of compact sources
             (Table~3).}
    \label{spectrum}
\end{figure}

In Fig.~\ref{spectrum}, these data are plotted together with flux density
values from \citet{rbs79} (408~MHz and 1415~MHz) and \citet{kpu94}
(83~MHz and 111~MHz).  All data are summarized in Table~3. For the Effelsberg $\lambda$21\ cm map 
observed in 1987,
we obtained the same flux density of $\rm 42.4\pm 4.1\ $Jy for G65.2+5.7 that was obtained 
for the map published by \citet{rbs79}. We subtracted the flux densities of
point-like sources listed in Table~2 by assuming a spectral index
of $\alpha = -0.79$, which we decided on after caöculating the average by weighting the source 
spectral indices according to their flux densities. The linear fit to the
integrated flux densities of G65.2+5.7 as shown in Fig.~\ref{spectrum} yields a spectral
index of $\alpha = -0.58\pm0.07$.

\begin{table*}
\caption{Integrated flux densities of G65.2+5.7, before ($S_{\nu}$)
and after ($S_{\nu}cor$) subtraction of point-like sources within the boundary of SNR.}
\label{sources}
\begin{tabular}{r c r@{.}l@{}l r@{.}l@{}l l }
\hline\hline
$\nu$ &Resolution &\multicolumn{3}{c}{$S_{\nu}$}  &\multicolumn{3}{c}{$S_{\nu}$cor}  & Ref. \\
(MHz)  &(arcmin)  &\multicolumn{3}{c}{(Jy)}       &\multicolumn{3}{c}{(Jy)}        &   \\
\hline
83     &310$\times$240   &200 & 0$\pm$ &50    &149 & 2$\pm$ &50     &\citet{kpu94} \\
111    &310$\times$240   &175 & 0$\pm$ &40    &134 & 7$\pm$ &40     &\citet{kpu94}\\
408    &   37            &91  & 0$\pm$ &15    &76  & 6$\pm$ &15     &\citet{rbs79}\\
1415   &   9             &42  & 4$\pm$ &4.6   &37  & 0$\pm$ &4.6    &\citet{rbs79}\\
2639   &   4.4           &    &        &      &21  & 9$\pm$ &3.1    &this paper\\
4800   &   9.5           &    &        &      &16  & 8$\pm$ &1.8    &this paper\\
\hline
\end{tabular}\\
\end{table*}


\begin{figure*}[!htbp]
\centering
\includegraphics[width=0.25\textwidth, angle=-90]{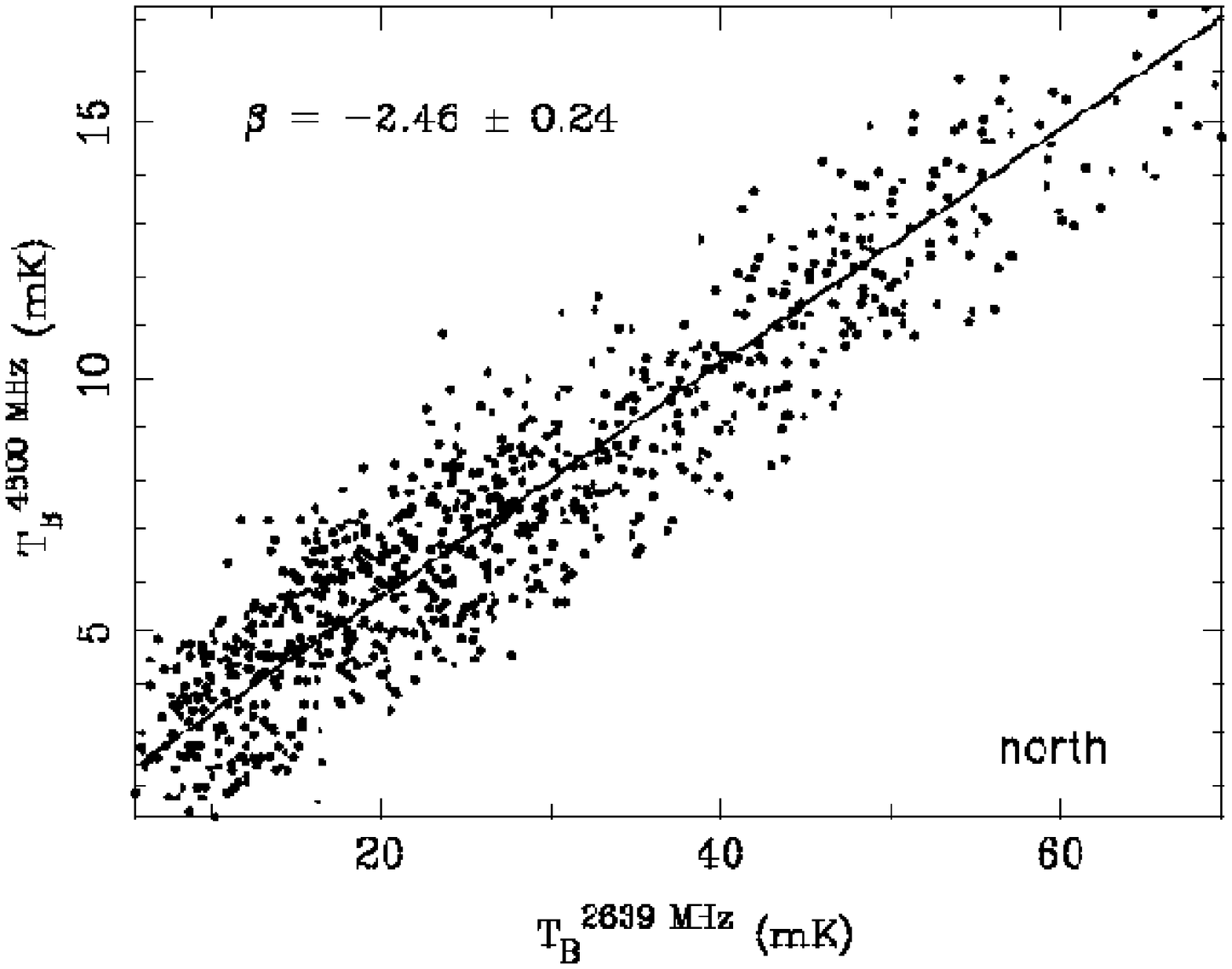}
\includegraphics[width=0.25\textwidth, angle=-90]{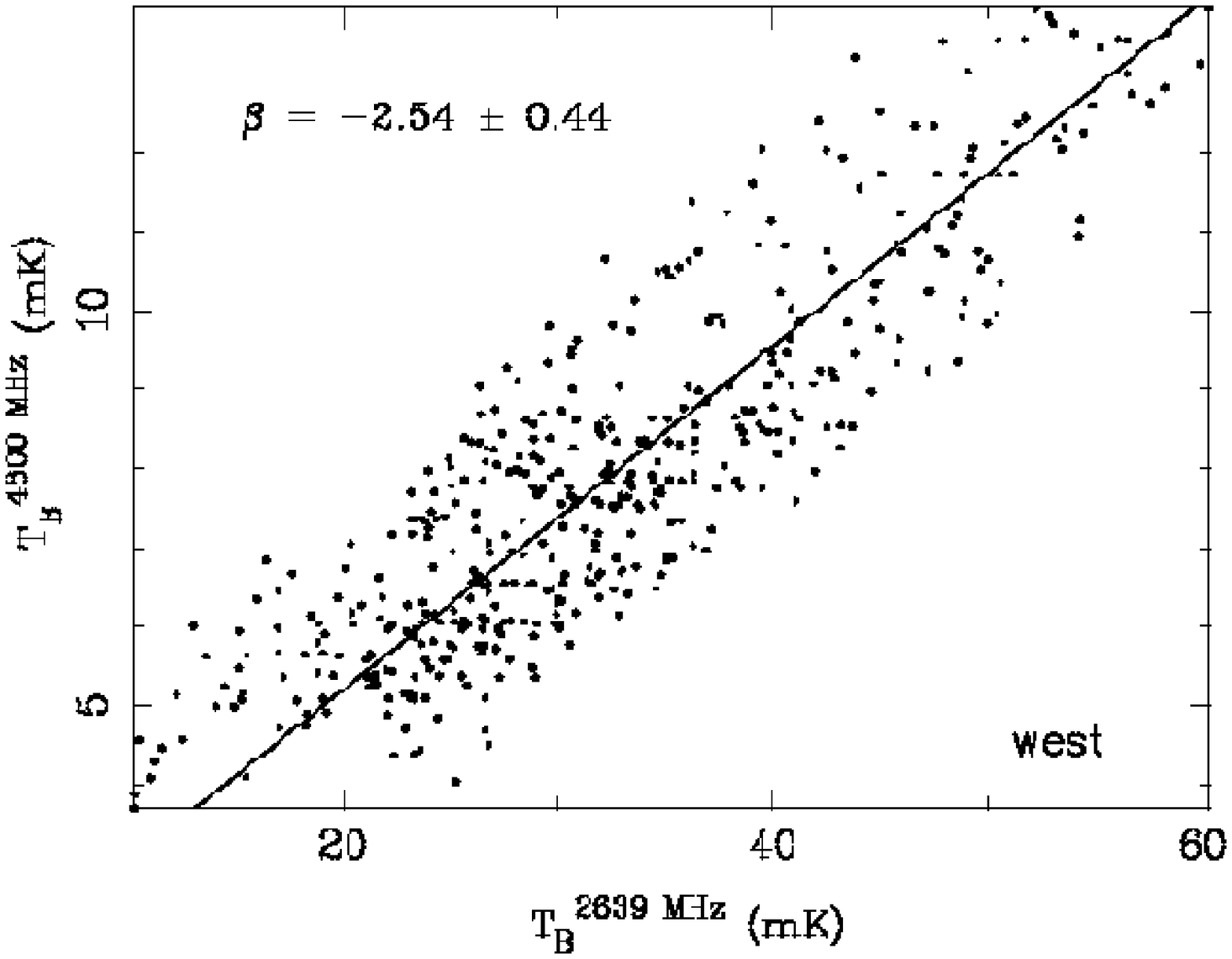}
\includegraphics[width=0.25\textwidth, angle=-90]{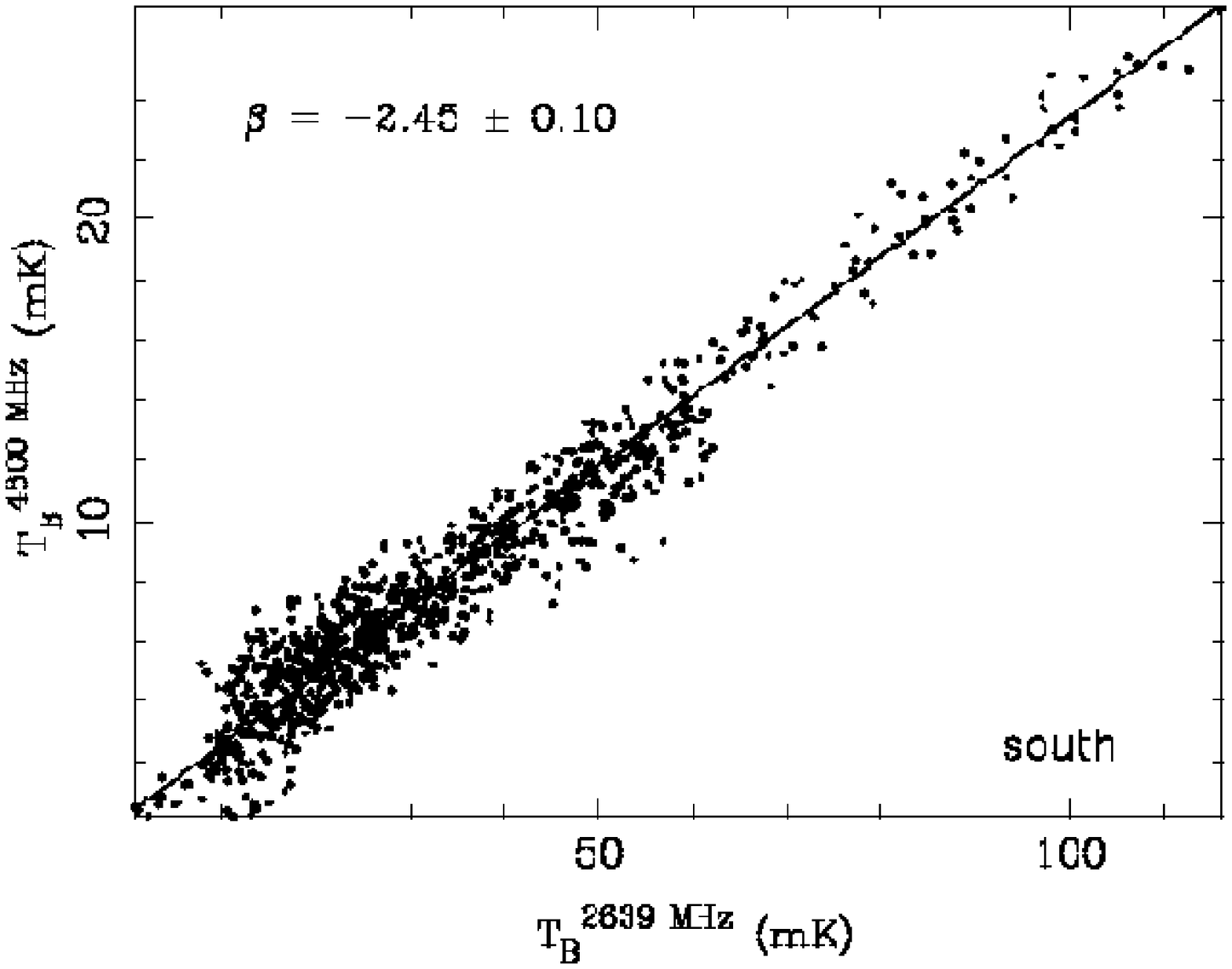}
    \caption{T-T plots for different shell sections of G65.2+5.7 between $\lambda$11\ cm and
             $\lambda$6\ cm. 
             }
    \label{tt}
\end{figure*}

\begin{figure}[!htbp]
\includegraphics[width=0.4\textwidth, angle=-90]{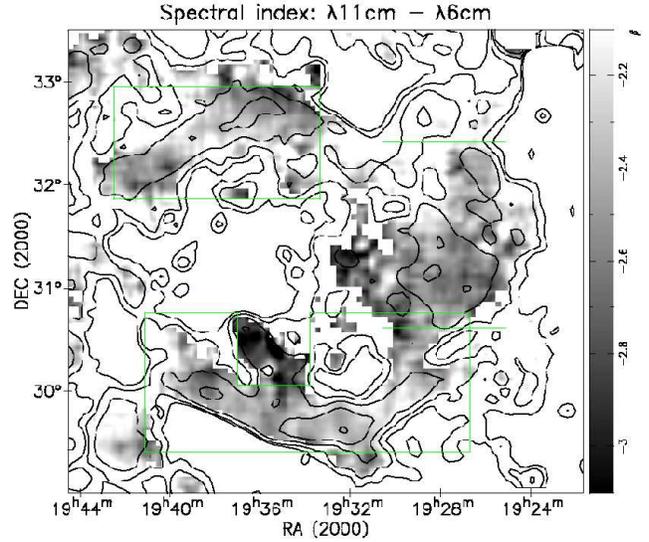}
    \caption{ 
             Spectral index map calculated between $\lambda$11\ cm and
             $\lambda$6\ cm at 10$\arcmin$ angular resolution for total intensities exceeding 12~mk~T$_{\rm B}$ 
             and 4~mk~T$_{\rm B}$ at $\lambda$11\ cm and $\lambda$6\ cm, respectively.
             The overlaid contours show total intensities at $\lambda$11\ cm starting at 12~mK~T$_{\mathrm B}$, 
             increasing in steps of $\rm 2^{n}\times 3\sigma_{I}$, where n=0,1,2,3$\cdots$, 
             and $\rm \sigma_{I} = 2.2~mk~T_{\rm B}$.
             The regions used for TT-plots in Fig.~\ref{tt} are indicated by boxes.}
    \label{index}
\end{figure}

\subsection{TT-plot spectral analysis}

The TT-plot method used for relative zero-level setting in Sect.~2.2 was also used to
investigate the spectrum of distinct emission structures in a way independent of
a consistent base-level setting of both maps.
Spectral index variations of structures within a source could then be recognized or
the integrated spectrum checked. 
We smoothed both maps to $10\arcmin$ and applied this method for three different sections of the G65.2+5.7 shell, 
which are marked in Fig.~\ref{index}.
The results are shown in Fig.~\ref{tt} for the $\lambda$11\ cm/$\lambda$6\ cm data.
A clear temperature-temperature relation is evident in all cases. The same TT-plot procedure
was used for the $\lambda$21\ cm/$\lambda$6\ cm data.
The temperature spectral index $\beta (= \alpha - 2)$ found from fitting the 
slope is $\beta_{11/6} = -2.45\pm0.10$ and $\beta_{21/6} = -2.61\pm0.24$ 
for the southern shell, $\beta_{11/6} = -2.46\pm0.24$ and $\beta_{21/6} = -2.65\pm0.27$
 for the  northeastern shell, and
$\beta_{11/6} = -2.54\pm0.44$ and $\beta_{21/6} = -2.71\pm0.45$ for the western shell.
The error in $\beta$ for the diffuse western part of the shell region is large, probably due to its weak
emission and confusion with weak unresolved background sources, which could not be subtracted.
All values agree with both each other and the spectral index of the integrated spectrum to within the errors.
However, we always obtain somewhat steeper spectra for the $\lambda$21\ cm/$\lambda$6\ cm data
than the $\lambda$11\ cm/$\lambda$6\ cm data, which reflects the slightly lower
integrated flux density at $\lambda$11\ cm relative to the fit shown in Fig.~\ref{spectrum}
and the marginally higher value we obtained at $\lambda$21\ cm.

\subsection{Spectral index map}

The Urumqi $\lambda$6\ cm and the Effelsberg $\lambda$11\ cm maps with point-like
sources removed and base-level corrected were both convolved to a common angular
resolution of $10\arcmin$. We calculated the spectral index for each pixel from
the brightness temperatures at the two frequencies.
We defined a lower intensity limit of 4~mK~T$_{\rm B}$ and  12~mK~ T$_{\rm B}$ for the
$\lambda$6\ cm and $\lambda$11\ cm map, respectively,
to achieve reasonable spectral indices without the influence of noise
and local distortions. 
We display the spectral index map between $\lambda$6\ cm and $\lambda$11\ cm in
Fig.~\ref{index}. Possible remaining variations in
the base-levels at $\lambda$11\ cm or $\lambda$6\ cm cause a systematic uncertainty in the
spectral indices of $\Delta \beta \sim 0.2$. The uncertainty is larger when
total intensities are lower.

The filament emerging from the southern shell towards the north has a somewhat steeper spectrum
than that of the southern shell. This is confirmed by $\beta_{11/6} = -2.75\pm0.28$ obtained from 
TT-plot, although this spectral index value is within the errors of those obtained for the other shell regions.

\citet{rbs79} found an indication of spectral flattening towards
the centre of G65.2+5.7 of about $\Delta \beta \sim 0.3$ when
comparing their Effelsberg $\lambda$21\ cm map with low-resolution data
at 408~MHz \citep{has74}.  This finding cannot be verified by the
present data. The emission in the central area is too faint
at higher frequencies to derive meaningful spectral indices in relation to
the base-level uncertainties.

\begin{figure}[!htbp]
\includegraphics[width=0.4\textwidth, angle=-90]{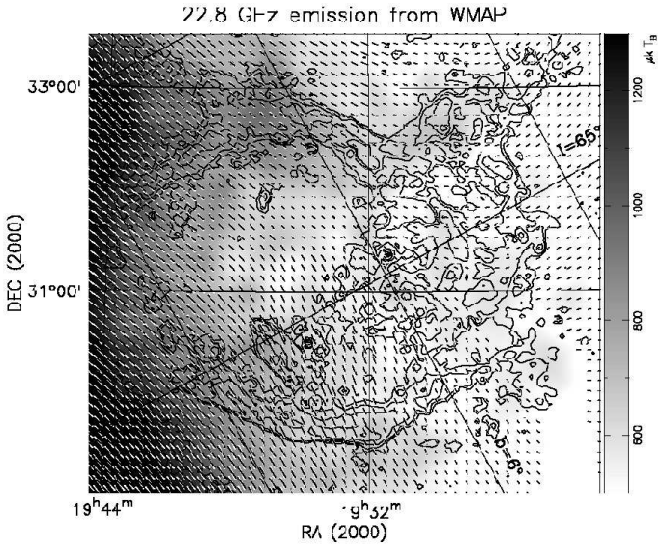}
    \caption{The WMAP 22.8~GHz map. 
             The total intensity emission (grey-scale) map is smoothed to an angular resolution of $2\degr$ with
             polarization B-vectors overlaid. Their length is proportional to the polarized intensity.
             The overlaid contours show total intensities at $\lambda$11\ cm at the same level as in Fig.~\ref{11cm}
             with compact sources removed.
             }
    \label{wmap}
\end{figure}

\subsection{The linear polarization properties of G65.2+5.7}

The $\lambda$6\ cm and $\lambda$11\ cm polarization maps shown in
the lower panels of Figs.~\ref{6cm} and \ref{11cm} indicate the similar polarization features at
the two frequencies. Four polarized patches appear at $\lambda$6\ cm, three
smaller and fainter ones located in the northern area of G65.2+5.7, and a single large patch
covering the southern shell and part of the inner area of the SNR. Apart from the enhanced
polarization along the SNR southern shell, most of the polarization
structures might originate from polarized diffuse interstellar emission
possibly mixed with polarized emission from the SNR of about similar
strength. We note low polarized emission along the northern shell at both
wavelengths. 

As shown in Fig.~\ref{21cmpi} the $\lambda$21\ cm polarization data do not indicate any 
polarized emission at all compared to G65.2+5.7, since neither the polarized intensity nor the 
polarization vectors change significantly in the direction of the SNR relative to its
surroundings. This indicates that most of the polarized emission seen in
this area at $\lambda$21\ cm originates from the foreground within the distance to
G65.2+5.7 of about 800~pc.

The large-scale emission components are missing in the two polarization maps
at $\lambda$6\ cm and $\lambda$11\ cm, because the end-points
of each scan in the individual U and Q maps were set to zero during data processing.
This limits the interpretation of polarization structures \citep{r06}, 
except for the strong polarized region along the southern shell at $\lambda$6\ cm.

\begin{figure}[!htbp]
\includegraphics[width=0.4\textwidth, angle=-90]{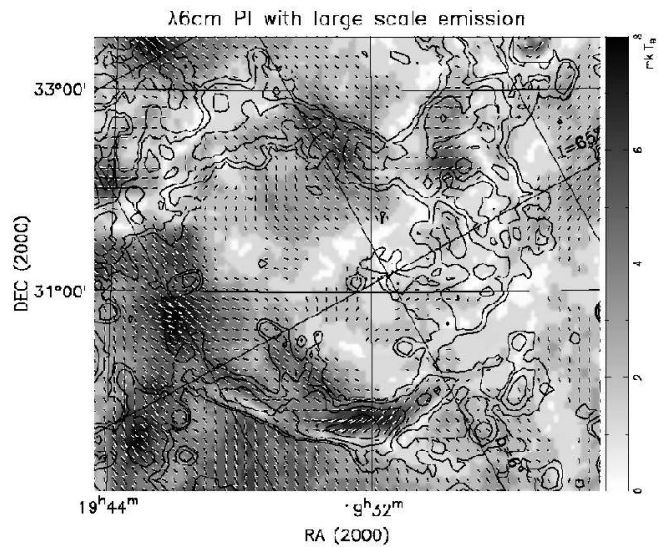}
    \caption{The $\lambda$6\ cm polarized intensity map (grey scale) with large-scale
             polarized emission restored from WMAP 22.8~GHz polarization data. B-field vectors are
             shown for the case of negligible Faraday rotation. Contours show $\lambda$6\ cm 
             intensities as in Fig.~\ref{6cm}.
             }
    \label{6cmwmap}
\end{figure}

Following \citet{shr07}, we used the WMAP polarization data at 22.8~GHz \citep{page07} 
to recover missing large-scale structures at $\lambda$6\ cm.  
There is no signature of G65.2+5.7 in the Stokes U and Q maps at 22.8~GHz. 
The convolved 22.8~GHz PI map shows 
a smooth intensity increase below Galactic latitudes of about $5\fdg5$. Above  $5\fdg5$,
the orientation of the polarization B-vectors is less regular (Fig.~\ref{wmap}). 
We convolved the WMAP 22.8~GHz U and Q maps corresponding to an angular resolution of $2\degr$ and 
scaled them to 4.8~GHz ($\lambda$6\ cm) assuming a temperature spectral index of
 $\beta=-2.65$, which was taken from the spectral index map of total intensities
between 408~MHz and 1420~MHz by \citet{rr88}. 
We assume the same spectral index for the Galactic diffuse polarized emission. 
The difference between the scaled WMAP data and the $\lambda$6\ cm maps, when convolved to the
same beam size of $2\degr$, results in an offset of  $\rm -0.4~mK~T_{\rm B}$ for Stokes U and
$\rm -2.3~mK~T_{\rm B}$ for Stokes Q, which is equivalent to $\rm 2.3~mK~T_{\rm B}$ of 
polarized intensity. The offsets were added to the original $\lambda$6\ cm U and Q data.
The polarized intensity map at $\lambda$6\ cm derived from the
restored data is shown in Fig.~\ref{6cmwmap}. 

We then compared the $\lambda$6\ cm polarization maps with and without 
absolute zero-level restoration. We found in general that the addition of missing
large-scale polarization did not change the morphology of the observed polarization
at $\lambda$6\ cm significantly. The polarized signal from the southern
shell of G65.2+5.7 was strong enough to remain almost unchanged when adding the
large-scale emission component.  Strong polarization emission of the diffuse
interstellar medium near to the Galactic plane  
in the south-eastern area of the map is unrelated to G65.2+5.7. 
However, we note some changes in areas of lower polarized emission. 
For instance, the polarized emisssion is more pronounced in
the direction of the extended HII-region LBN~150, also known as Sh2-96, ($\rm 19^{h}
28.7^{m}, +32\degr 41\arcmin$, J2000) in Fig.~\ref{6cmwmap}, which
itself does not emit polarized emission, but may act as a Faraday screen by rotating
polarized emission from larger distances. 

The procedure for recovering missing large-scale emission by using the WMAP polarization data,
needs to be modified in case the Faraday rotation is too large or becomes significant
at longer wavelengths because of the $\lambda^{2}$-dependence of the polarization angle for 
a given RM. Therefore, it seems questionable that the $\lambda$11\ cm U and Q maps should by corrected 
in the same way as the $\lambda$6\ cm polarization data. 
The level of Galactic PI at $\lambda$11\ cm, however, is calculated to be about 
$\rm 11.5~mK~T_{\rm B}$. This large-scale polarized signal needs to be compared with the observed 
polarized signals at $\lambda$11\ cm,
which are only three times stronger on the inner side of the southern shell and only twice as strong
for the average polarization of the entire SNR. 
We conclude that the missing large-scale polarization is a serious contamination of the
$\lambda$11\ cm data and any RMs calculated from the $\lambda$11\ cm and 
the $\lambda$6\ cm data will by affected by an unknown amount by that effect.

\subsection{The northern area of G65.2+5.7}

The polarized emission seen in the northern area of the SNR is much weaker than in the southern
area and not very clearly related to the SNR shell. Along the northern shell,
depolarization at both $\lambda$11\ cm and $\lambda$6\ cm is evident, which may
have various causes. Firstly, background polarization is depolarized by small-scale
magneto-ionic fluctuations in the shell; secondly,
polarized emission from the SNR shell has a different orientation and cancels, at least partly, the
background emission; or, thirdly, Faraday rotation in the shell rotates background polarization
in such a way that it cancels the foreground polarization.
A combination of the different scenarios is certainly possible. 
We note that \citet{mbp02} measured the highest thermal electron densities for 
the optical filaments in the northern shell, which may explain the high depolarization. 
The low percentage polarization in the northern area, compared to that of the southern shell, may 
indicate a less regular magnetic field. We conclude that the magnetic field
properties along the northern shell cannot be determined reliably from the data available to us.

\begin{figure}
\centering
\includegraphics[width=0.15\textwidth, angle=-90]{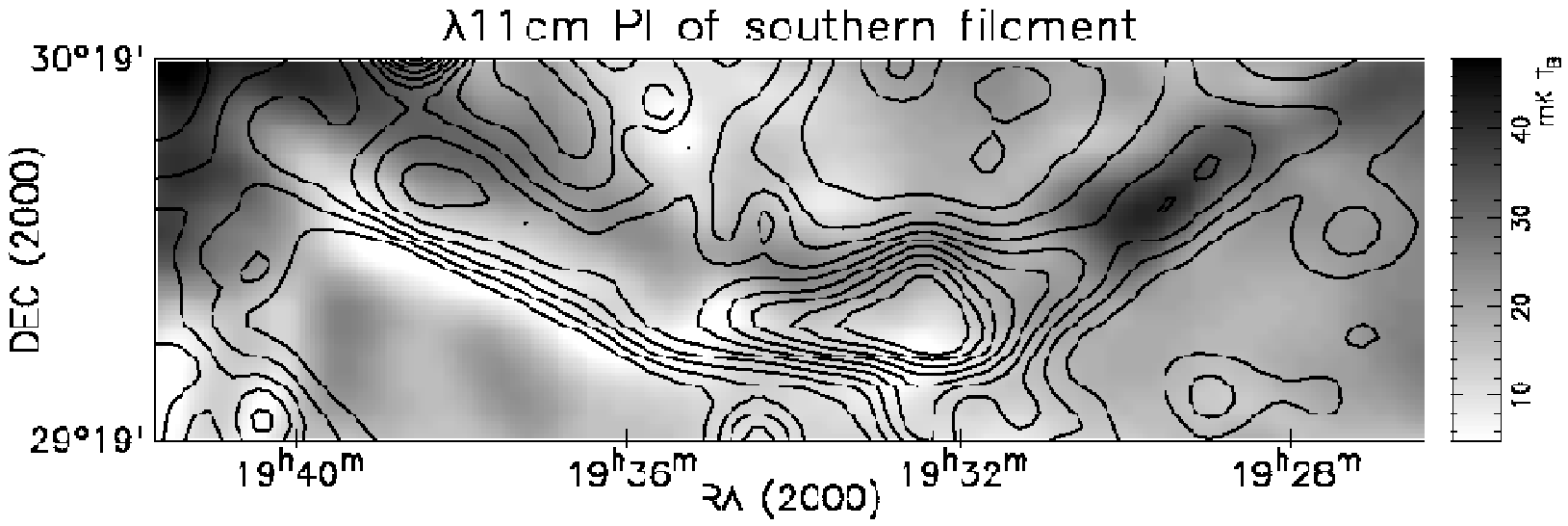}
\includegraphics[width=0.15\textwidth, angle=-90]{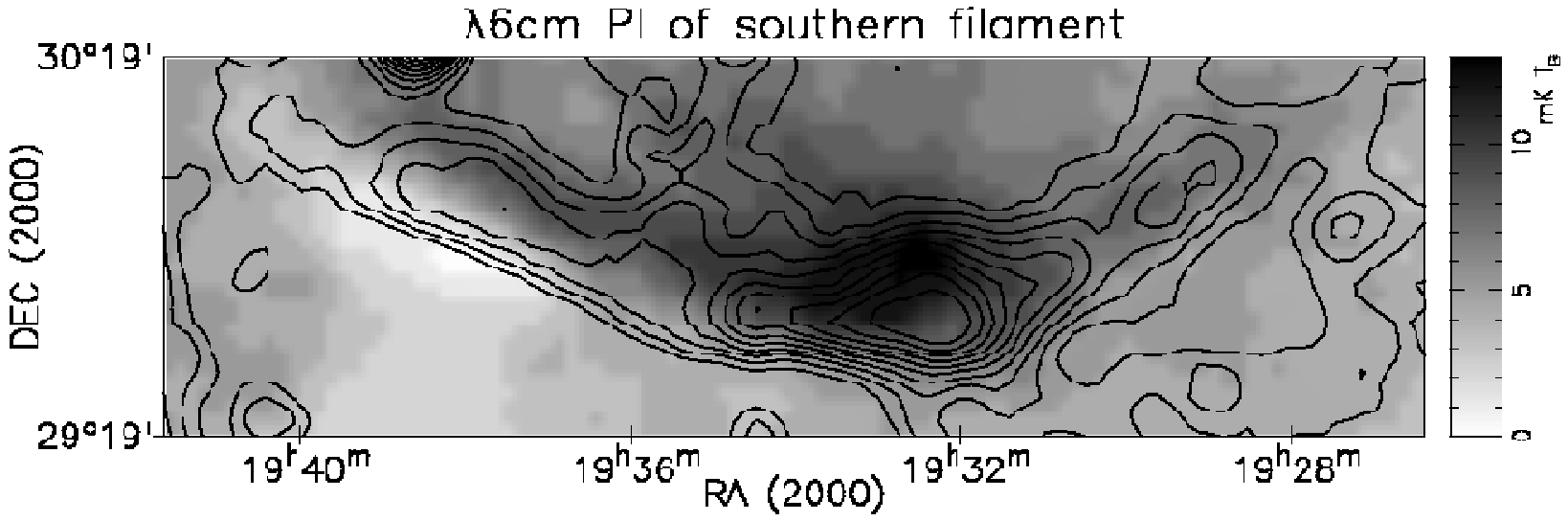}
    \caption{The observed polarized intensities in grey scales for the southern shell 
             at $\lambda$11\ cm and $\lambda$6\ cm overlaid with
             contours of corresponding total intensities.
             Both maps are smoothed to $10\arcmin$ angular resolution.
             }\label{south}
\end{figure}

\subsection{The southern shell of G65.2+5.7}

As shown in Fig.~\ref{south} the polarized emission at $\lambda$6\ cm reaches its maximum at 
the southern shell. The polarization percentage of the southern shell locally is as high 
as 54\% at $\lambda$6\ cm. Lower polarized emission at $\lambda$11\ cm
at the same position indicates strong depolarization at this wavelength. 
Inside the shell the polarization percentage at $\lambda$11\ cm is about 38\%.

As expected for an evolved SNR, a well ordered magnetic field is detected in the shell.
Rotating the polarization E-vectors shown in Fig.~\ref{6cm} by $90\degr$ results in an observed
magnetic field direction that is almost tangential to the shell (Fig.~\ref{6cmwmap}). 
Thus, the magnetic field intrinsic to the shell is tangential when the Faraday rotation 
is low. If we accept the deviations (of about $15\degr$) from tangential orientations as
the result of intrinsic plus foreground RMs, then the total RM should not exceed about $\rm 70~rad~m^{-2}$.
 
We estimated the contribution of RM from the foreground interstellar medium. The NE2001 model 
of \citet{cl02} predicts a dispersion measure of $\rm DM~=~9~pc~cm^{-3}$ in the direction of G65.2+5.7. 
The magnetic field strength along the light-of-sight,
$B_{||}$, is about $\sim 2\mu$G, which is a typical value
for the local regular magnetic field \citep{hml06}. Then RM is calculated by
RM~=~0.81~B$_{||}$~DM resulting in about $\rm -14~rad~m^{-2}$.
This Galactic foreground RM has little
effect (only a few degrees) on the polarization vector orientation measured at $\lambda$6\ cm.

\begin{figure*}[!htbp]
\includegraphics[width=1.0\textwidth, angle=0]{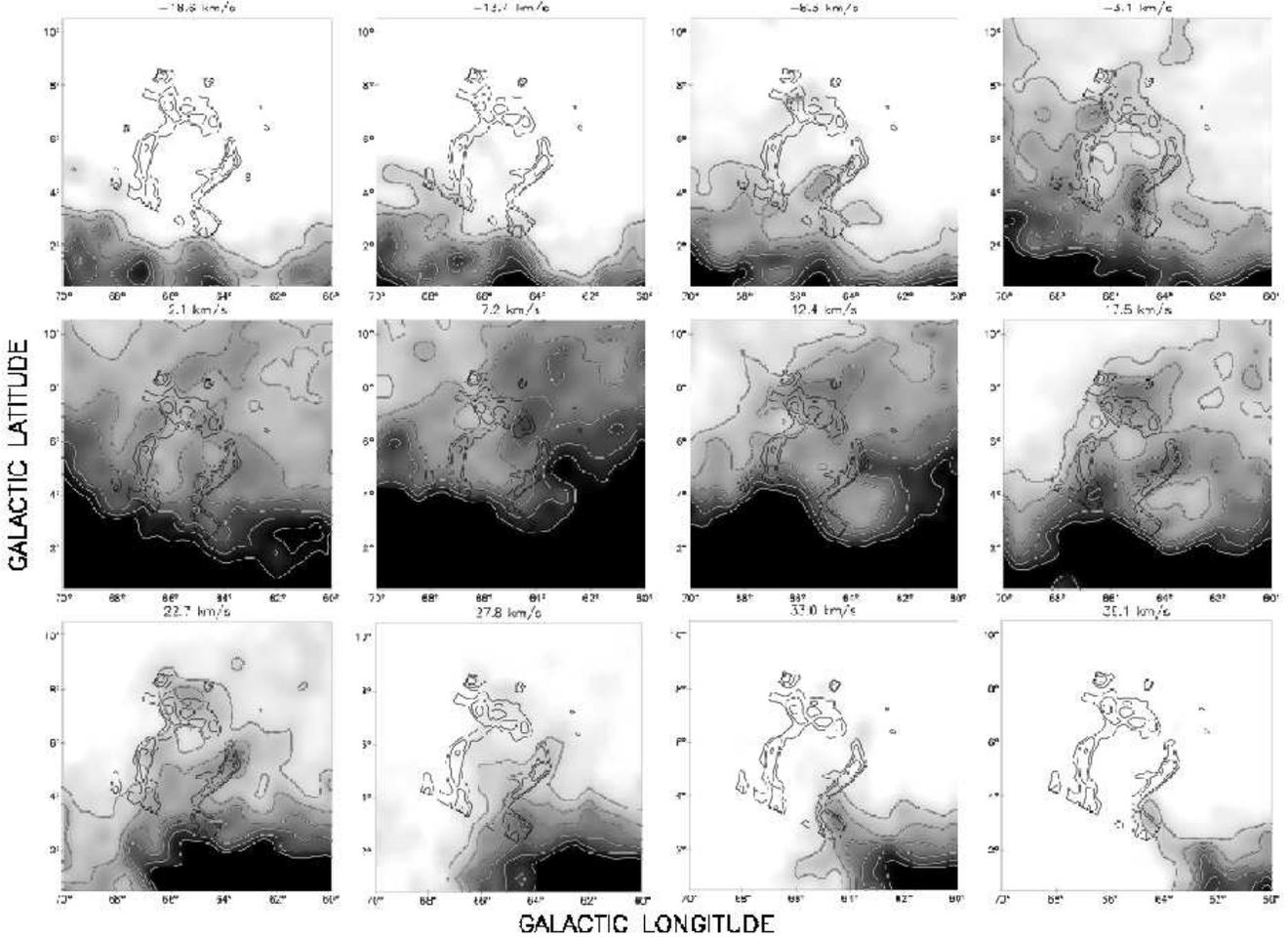}
    \caption{The $\lambda$21\ cm HI-intensities in the area of G65.2+5.7 for the velocity range 
             from $\rm -20~km~s^{-1}$ to 40~km~s$^{-1}$,
             averaged over five consecutive channels ($\sim$5~km~s$^{-1}$). 
             The central LSR velocities are indicated above each panel.
             The grey-scale ranges between 40 and 300~K~km~s$^{-1}$.
             White contours are shown from 80~K~km~s$^{-1}$ (except for the 2.1~km~s$^{-1}$ map, which starts at
              100~k~km~s$^{-1}$) and increases in steps of 40~K~km~s$^{-1}$. The  
             $\lambda$6\ cm total intensities smoothed to $18\arcmin$ are superimposed as contours in black.
  The northern shell is located at about $67\degr$ longitude and the southern shell at about $64\degr$. }
    \label{channel}
\end{figure*}

When rotating the E-vectors at $\lambda$11\ cm in the same way as at $\lambda$6\ cm,
magnetic directions that differ up to $80\degr$ from a tangential orientation direction imply
values of RM of the order of 150~rad~m$^{-2}$ for the $\lambda$11\ cm data to agree with a 
tangential magnetic field direction. This large RM disagrees with the $\lambda$6\ cm data.
However, as stated above, the observed level of the $\lambda$11\ cm polarized emission is too low 
to be unaffected by unrelated Galactic large-scale emission. This needs to be taken 
into account before calculating any RM for G65.2+5.7, but we have insufficient information
to do this.
Clearly, we require observations at wavelengths shorter than $\lambda$6\ cm of sufficiently 
high angular resolution
to calculate the RM properties of G65.2+5.7 in combination with our $\lambda$6\ cm data 
and thus contain the suggested tangential magnetic field orientation along its southern shell.

We can estimate the magnetic field strength in the southern shell by assuming
energy equipartition between the magnetic field and electrons and protons
in the SNR-shell using the formula by \citet{fr04}, where the radius of a source
is replaced by the volume V to deal with a shell section

\begin{equation}
B_{\min}  \approx 175 \cdot \Phi\, \mbox{}^{-2/7} \cdot V(pc^{3})\, \mbox{}^{-2/7} \cdot d(kpc)\, \mbox{}^{4/7} \cdot S_{1GHz}(Jy)\, \mbox{}^{2/7}
\end{equation}

We extrapolated the integrated $\lambda$6\ cm flux density without compact sources for the
southern shell section centred on $\rm \alpha_{2000}=19^{h}33^{m}$,
$\delta_{2000}=30\degr$ ($4\degr$ in $\alpha$, $1\degr$ in $\delta$)
of $4.0$~Jy to $8.8$~Jy at 1~GHz for the error-weighted mean spectral index from TT-plots corresponding to
 $\alpha =-0.50$. We estimated the thickness of the southern shell 
from our highest angular resolution data at $\lambda$11\ cm to be
about 10\% of the radius, or about 2.3~pc. 
For a distance of 0.8~kpc, we calculated a volume of the southern shell region of about 
$\rm 6500~pc^{3}$. The estimated equipartition magnetic field $\rm B_{min}$ is about 
 $23~\mu$G for a filling factor $\Phi$ = 1, although this is a lower limit. In case
of a highly filamentary shell, such as G65.2+5.7, a filling 
factor of $\Phi$ = 0.1 may be more appropriate, which would increase $\rm B_{min}$ to $45~\mu$G.

Another approach to estimating equipartition magnetic fields in SNRs was described by \citet{f05}.
This method results in a magnetic field strength of  $\rm B \sim 36~\mu$G for
$\alpha =-0.50$ over a frequency range of $\rm \nu_{min,max}=0.1-100~$GHz, 
a line-of-sight extension of the radio-emitting region of 50~pc, and a radio intensity of the southern shell 
of $\rm I_{\nu}=0.2\times 10^{-18} erg s^{-1} cm^{-2} Hz^{-1} sr^{-1} $. 
We conclude that the magnetic field of the southern shell has a strength of between $\rm 20~and~50~\mu$G. 

\section{Discussion}


The radio properties of G65.2+5.7 show a number of similarities with another bilateral
SNR G156.2+5.7 \citep{rfa92, xhs07}, which also has the lowest surface brightness
of all known SNRs with a surface brightness of 
$\rm \Sigma_{1 GHz} = 5.8\times 10^{-23}$ W$\rm m^{-2} Hz^{-1} sr^{-1}$
at l~GHz. The surface brightness of G65.2+5.7 is about $\rm \Sigma_{1 GHz} = 
1.3\times 10^{-22}$ W$\rm m^{-2} Hz^{-1} sr^{-1}$, just about twice that of G156.2+5.7.
Both SNRs are located outside the Galactic plane and at
a similar distance of 800~pc or 1~kpc, although the shape of G65.2.+5.7 is more elliptical and
its size is about 50\% larger, possibly indicating that it is a more evolved object. Both 
are very bright in X-rays.
The symmetry axis of both objects shows a large inclination relative to the Galactic
plane, which has been taken as evidence that the ambient magnetic field has a similar inclination.
The WMAP 22.8~GHz polarization data at an angular resolution
of $2\degr$ in Fig.~\ref{wmap} indicate that the magnetic field is aligned almost parallel to the Galactic
plane up to about $6\fdg5$ latitude. Above $6\fdg5$ latitude, the direction is less clearly defined.
Of course, the WMAP data show the integrated polarization along the line-of-sight and 
local deviations in the magnetic field might be possible.

Some differences between G65.2+5.7 and G156.2+5.7 exist in their
polarization properties. At $\lambda$6\ cm and $\lambda$11\ cm G156.2+5.7 has a higher fractional
polarization. Depolarization along the outer shell of G65.2+5.7 seems to 
support the classification of a ``thermal composite'', based on X-ray
data, where a dense outer shell causes absorption of X-ray emission and  
strong depolarization \citep{skp04}.

\citet{obr07} presented simulations of the bilateral morphology of SNRs caused by density
gradients in the interstellar medium or a gradient in the ambient magnetic
field perpendicular to the radio shell. Smooth gradients are probably a simplification of
the true conditions. However, we note that the intensity differences between
the two arcs of the shell of G65.2+5.7  
seem to agree with some of the simulations by \citet{obr07} (their Fig.7, panels A and B),
in which quasi-parallel particle injection was assumed and the viewing angle was 
perpendicular to both $\langle B \rangle$ and the gradients of either density or 
magnetic field strength. Slanting similar radio arcs were produced
if the gradient was perpendicular to $\langle B \rangle$. In the quasi-perpendicular case, \citet{obr07} 
found a bilateral shape when the gradient of the ambient $\langle B \rangle$ or density was parallel 
to $\langle B \rangle$.
The symmetry axis of the SNR is always aligned with the gradient of density or the magnetic field.
As indicated by the WMAP total intensity map (Fig.~\ref{wmap}) or 
$\rm H_{\alpha}$ data (Fig.~\ref{multibands}, right panel), thermal emission is evident
along the northern and southern shell, and more smooth in the north than in the south.
In addition to the general density gradient in the direction of Galactic latitude, this indicates a 
density difference in the direction of longitude, which results in compression differences in the 
interacting SNR shell and to a brightness difference as observed.

\subsection{Analysis of HI-observations}

\citet{kbh05} published the Leiden/Argentina/Bonn (LAB) $\lambda$21\ cm survey of neutral hydrogen, 
which covers the velocity range from $\rm -450~km~s^{-1}$
to $\rm +400~km~s^{-1}$ with a velocity resolution of $\rm 1.3~km~s^{-1}$. The HPBW of the survey was $0\fdg6$,
spectra were taken every $0\fdg5$ and the r.m.s. brightness-temperature noise was about $\rm 0.07-0.09~K$. 
  
A fully sampled HI-survey was also carried out with the DRAO 26-m telescope by \cite{hig05}.
We found no significant difference between these two surveys, except that the DRAO HI-survey does
not cover the very northern part of the SNR. Our HI-mass calculation is thus based on the LAB survey.

We extracted the HI-data of the G65.2+5.7 region from the LAB survey data-cube in the velocity range
between $\rm -20~km~s^{-1}$ and $\rm 40~km~s^{-1}$. 
The 0.8~kpc distance corresponds to a velocity range $\rm 0-10~km~s^{-1}$ according
to the Galactic rotation model given by \citet{f89} of $\rm R_0 = 8.5~kpc$ and $\rm \Theta_0 = 220~km~s^{-1}$. 
Figure~\ref{channel} displays the HI distribution after averaging five
consecutive channels ($\rm \sim~5~km~s^{-1}$) within the velocity interval from $-20$ to $\rm 40~km~s^{-1}$.
The central velocity of each image is indicated at the top. 
Contours of $\lambda$6\ cm total intensity (smoothed to $18\arcmin$) are superimposed to indicate the region
of G65.2+5.7. 

Although HI-fluctuations are visible across the entire field,
we note signs of possible SNR interaction with the HI-gas in the images centered on $\rm v = -3.1~km~s^{-1}$, 
$\rm 2.1~km~s^{-1}$, and $\rm v = 7.2~km~s^{-1}$.
The northern shell correlates with the lower velocity maps, while the southern shell correlates with 
the $\rm 7.2~km~s^{-1}$ cloud. This may indicate that the northern shell is on the near side, or expands towards us,
while the southern shell is on the far side or expands away from us.   

\begin{figure}[!htbp]
\includegraphics[width=0.45\textwidth, angle=-90]{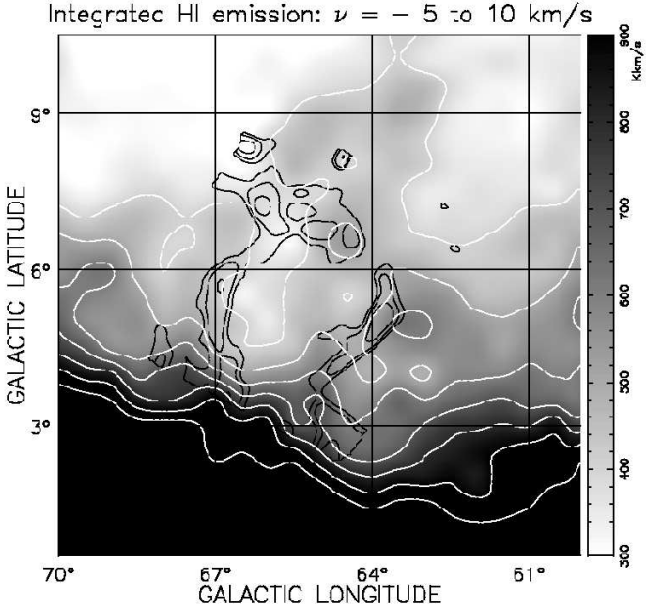}
    \caption{Integrated HI-intensities for the velocity range from $\rm -5~km~s^{-1}$ to 10~km~s$^{-1}$ superimposed on
             with $\lambda$6\ cm contours smoothed to $18\arcmin$. 
             White contours for HI intensities are shown from 400~K~km~s$^{-1}$ and run in steps of 100~K~km~s$^{-1}$.}
    \label{HI}
\end{figure}

In Fig.~\ref{HI},
we present the result for the integrated HI column densities over the velocity range $-5$ to $\rm 10~km~s^{-1}$. 
We integrated HI column densities over the velocity range $\rm -5~to~10~km~s^{-1}$, and subtracted
the large-scale diffuse emission 
using the 'unsharp-masking' procedure described by \citet{sr79}. 
A weak HI~shell with an angular diameter of about 4$\degr$ is visible at the outer periphery
of the radio continuum shell of SNR G65.2+5.7.

If a physical association exists, we would calculate the column density for the shell to be 
$\rm 2~\times~10^{20}~cm^{-2}$. Assuming a thickness of the shell of 20\% for a 2$\degr$ radius, 
this corresponds to a swept-up mass of about 1400~M$\odot$. 
The average density in the HI-shell is then about $\rm 1.2~cm^{-3}$.
If this mass of neutral gas was originally uniformly distributed within a sphere of radius 2$\degr$,  
the pre-explosion ambient density would have been $\rm 0.6~cm^{-3}$, 
which is comparable to the value of $\rm 0.4~cm^{-3}$ estimated by \citet{m79} for 
the X-ray emitting region.
These densities are reasonable for a distance of $\sim$80~pc away from the Galactic plane.

\subsection{Comparison with observations from other bands}

\begin{figure}[!htbp]
\includegraphics[width=0.49\textwidth, angle=0]{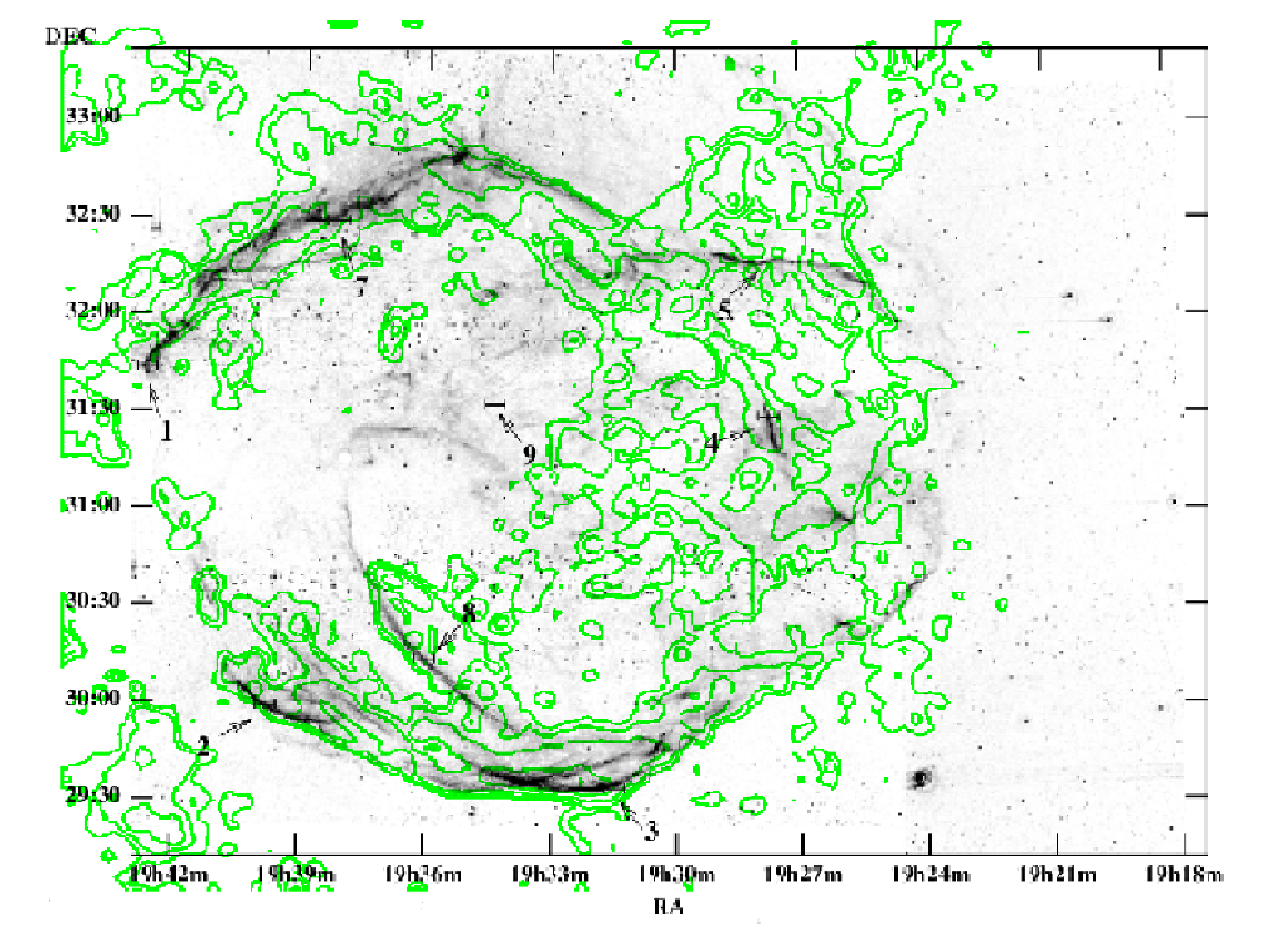}
    \caption{Superposition of an [O \MakeUppercase{\romannumeral 3}]~image \citep{bml04} 
             with $\lambda$11\ cm total intensity contours. The $\lambda$11\ cm map has 
             the same levels as shown in Fig.~\ref{11cm} but the compact sources were subtracted.
             }
    \label{optical}
\end{figure}

We show contours of $\lambda$11\ cm total intensities of G65.2+5.7 superposed on an 
[O \MakeUppercase{\romannumeral 3}] emission image by \citet{bml04} in Fig.~\ref{optical}.
The radio filaments are highly correlated with the 
[O \MakeUppercase{\romannumeral 3}] filaments, which trace the fast SNR shock, except for the northwestern more diffuse part of the SNR.
In this area, the SNR shock front seems to be distorted resulting in a less compressed
magnetic field and a reduced efficiency in particle acceleration.
We note that in some regions enhanced radio filamentary emission is seen without any
corresponding optical emission. The magnetic fields might be stronger
and/or the temperatures/densities lower in these bright radio filaments.
This seems to be consistent with the measured incomplete cooling regions of G65.2+5.7 \citep{mbp02},
which means that at least parts of the SNR are still in an adiabatic phase.

The G65.2+5.7 image ($\rm 0.14-0.284~keV$) from the ROSAT
soft X-ray all-sky survey \citep{sn97} is shown in Fig.~\ref{multibands} (left panel)
with contours of $\lambda$11\ cm total intensity superimposed.
The radio contours trace the outer SNR shock and clearly define an outer 
boundary of the X-ray emission. With its centrally filled X-ray thermal emission, G65.2+5.7
has been classified as a ``thermal composite'' SNR. 
\citet{skp04} reviewed the evolution of ``thermal composite'' SNRs: A dense
outer shell develops when the SNR enters the cooling phase, while the
strong centrally peaked thermal X-ray emission may be explained by
thermal conduction behind radiative shocks \citep{c99} and/or by the 
evaporation of interstellar clumps \citep{wl91}.
More studies are needed to develop a detailed scenario of the evolution.

We show $\lambda$11\ cm contours 
superposed on  a $\rm H_{\alpha}$ image of G65.2+5.7 \citep{f03} in
Fig.~\ref{multibands} (right panel). A good correlation exists between the radio
structures and enhanced $\rm H_{\alpha}$ emission, in particular along the southern shell.
We estimate the thermal electron density within the southern shell using the
observed emission measure (EM); this is defined as the integral of the
square of electron density of the source along the line-of-sight {\it l}, $\rm EM = n_e^2$ $l$,
and related to the $\rm H_{\alpha}$ intensity by

{ \begin{equation}\label{em}
EM=2.75T_4^{0.9}I_{\rm H\alpha}\exp\left(\tau \right),
\end{equation}
where $EM$ is in units of pc~cm$^{-6}$, $I_{\rm H\alpha}$ is the
H$\alpha$ intensity in Rayleigh, and the optical depth $\tau$ could
be written in units of the product of $\rm E(B-V)$ magnitudes of reddening 
and 2.44 according to \citet{f03}.

Assuming an electron temperature of 10000~K for the thermal gas and
given an $\rm H_{\alpha}$ intensity of 18~Rayleigh for the southern filament
by filtering out the diffuse emission,
we obtained an EM of about 145~pc~cm$^{-6}$ for a reddening $\rm E(B-V)$ of 0.44 \citep{sfd98}.
We then obtained an electron density $\rm n_e$ of between 38~cm$^{-3}$ and 22~cm$^{-3}$
for a thickness of between 0.1~pc and 0.3~pc. 
For the equipartition magnetic fields of $20~\mu$G to $50~\mu$G (filling factor 0.1), we may then 
calculate RMs ranging between 70~rad~m$^{-2}$ and 245~rad~m$^{-2}$
for the range of filament thicknesses and magnetic fields. 
We recall that for the case of a tangential magnetic field, the $\lambda$6\ cm
polarization angles constraint is about RM $\le \rm 70~rad~m^{-2}$. 

\begin{figure*}[!htbp]
\includegraphics[width=0.42\textwidth, angle=-90]{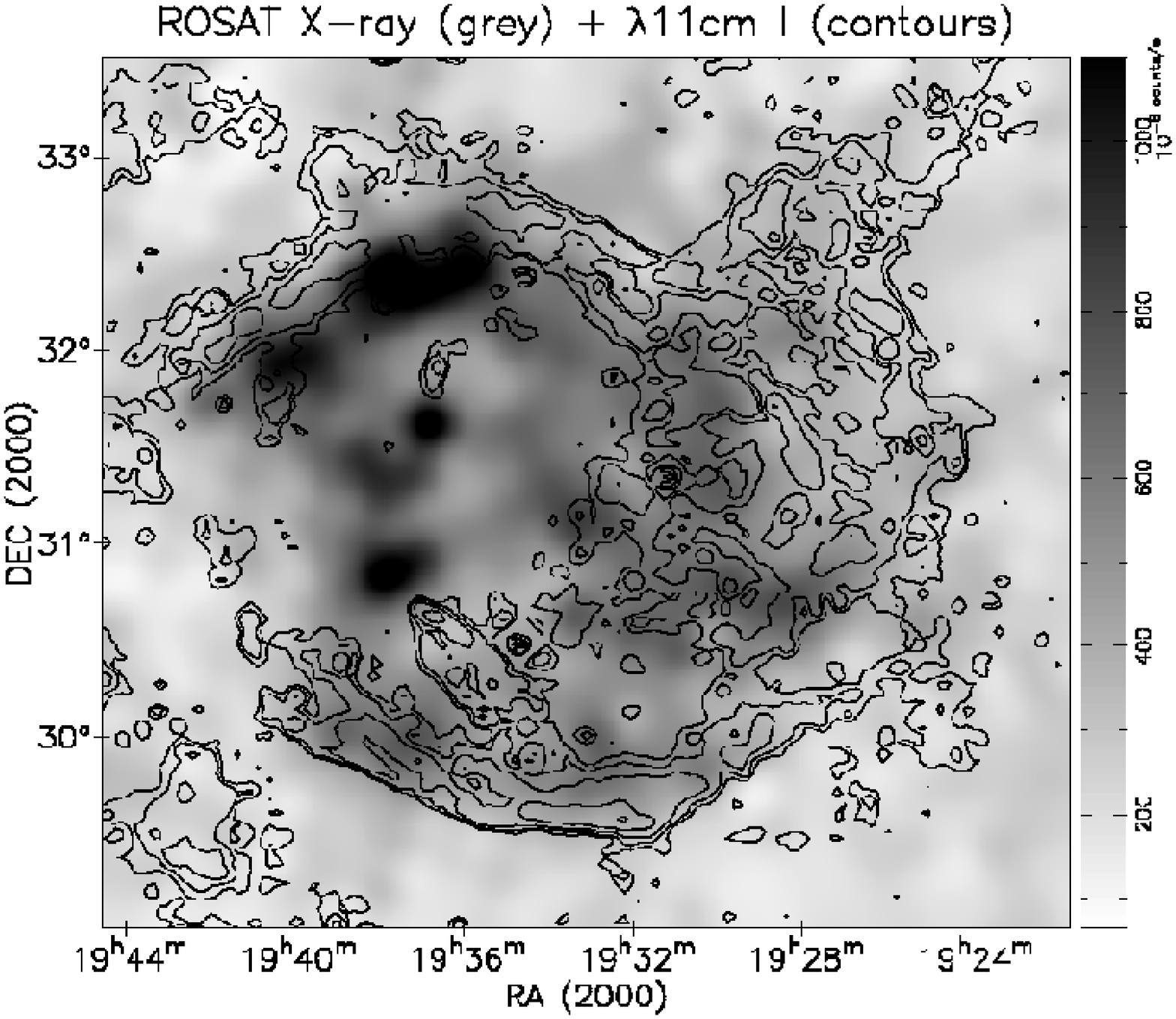}
\includegraphics[width=0.42\textwidth, angle=-90]{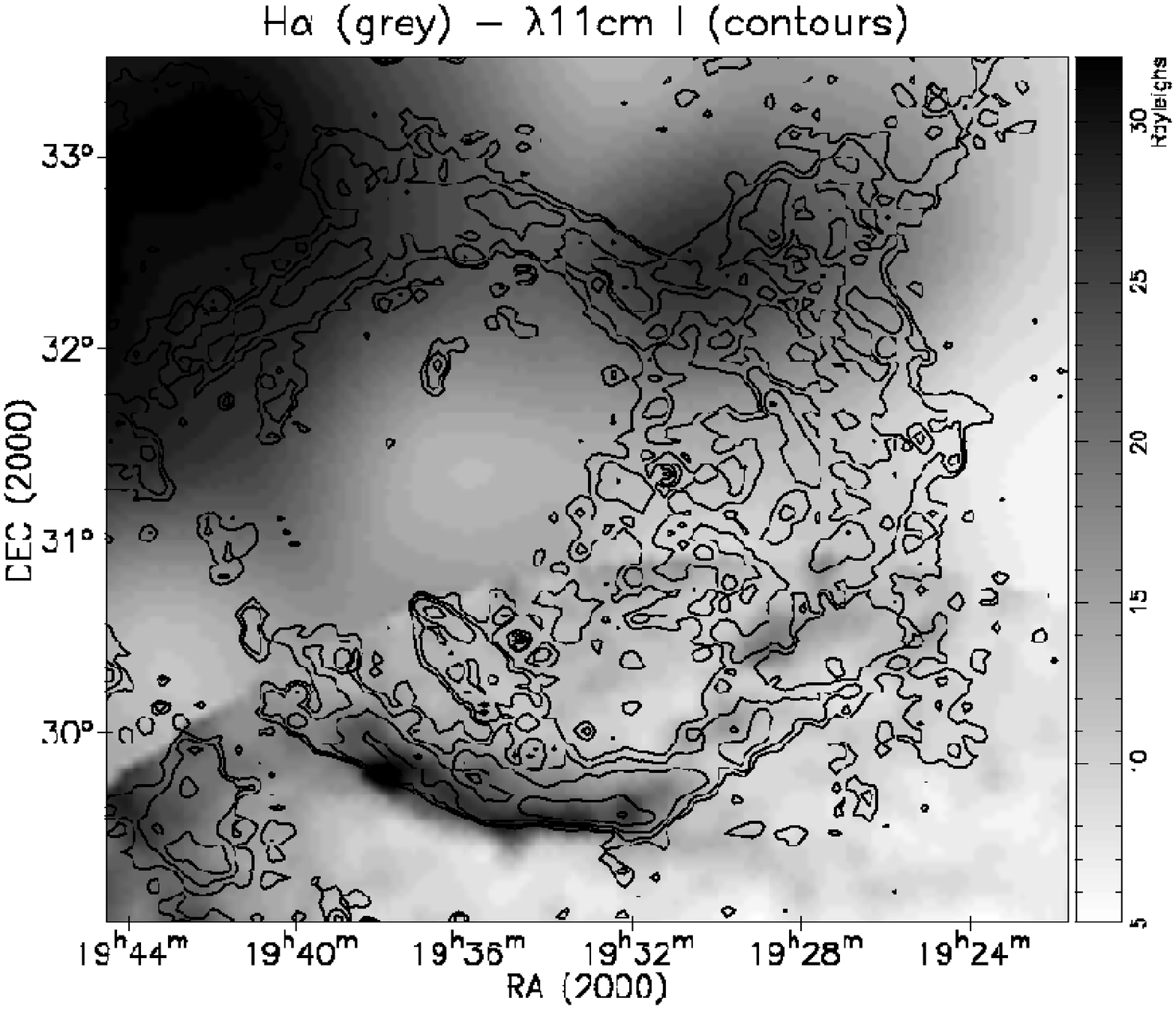}
    \caption{Superimposed images of SNR G65.2+5.7. Left:
             ROSAT X-ray emission \citep{sn97} (grey-scale) smoothed to $10\
arcmin$ angular resolution overlaid
             with $\lambda$11\ cm total intensity contours .
             Right: $\rm H_{\alpha}$ map taken from \citet{f03} (grey-scale)
 overlaid with $\lambda$11\ cm contours.
             Contour levels are as in Fig.~\ref{11cm}. Note that compact sources were removed from the
$\lambda$11\ cm data shown here.
             }
    \label{multibands}
\end{figure*}

We also investigated the distribution of CO-emission from the ``Composite CO-Survey'' \citep{duc87}
for the SNR region, but found no evidence of any associated emission.

\section{Summary}

We have presented high sensitivity maps of the SNR G65.2+5.7 for  total and polarized
intensities at $\lambda$6\ cm and $\lambda$11\ cm. The spectral index
of the integrated flux densities was found to be $\alpha=-0.58\pm0.07$, which 
is consistent with the value obtained by \citet{rbs79}.
The distribution of spectral indices shows some variation, although
within the error margins of the spectral index obtained for the integrated emission.

Strong polarized emission is observed from the southern filament of G65.2+5.7, where a
high percentage polarization around 50\% at $\lambda$6\ cm indicates the presence of a strong 
regular magnetic field component.
Clear depolarization along its outer rim is seen at $\lambda$11\ cm.
The polarized emission at $\lambda$11\ cm and
$\lambda$6\ cm are affected by a different amount of depolarization as well as confusion from
diffuse Galactic polarized emission.

G65.2+5.7 is a faint large-diameter shell-type SNR with an exceptional low surface
brightness in the radio range \citep{rbs79}, but otherwise no unusual properties, e.g. no
spectral bend as SNR S147 \citep{xfrh08}.  Except for possibly a few areas,
G65.2+5.7 has almost entered the cooling phase.
The strong depolarization along the outer SNR shell seems to support the classification
of G65.2+5.7 as a rare example of a ``thermal composite'' SNR based on X-ray data.

\begin{acknowledgements}

We thank Dr. XiaoHui Sun for useful
discussions and Dr. Patricia Reich for reading and comments on the
manuscript. We are very grateful to the referee Dr. Tyler Foster for
constructive comments, which clearly improved the manuscript.
The $\lambda$6\ cm data were obtained with a receiver system from
the MPIfR mounted at the Nanshan 25m telesope at the Urumqi
Observatory of NAOC. 
The nice receiver was constructed by Mr. Otmar Lochner of MPIfR,
and well-maintained by Mr. M. Z. Chen and J. Ma of Urumqi observatory.
The $\lambda$11\ cm observations are based on
observations with the 100-m telescope of the MPIfR (Max-Planck-Institut
f\"ur Radioastronomie) at Effelsberg. This
research work was supported by the National Natural Science Foundation of
China (10773016, 10833003 and 10821061), and the Partner group of the MPIfR at
NAOC in the frame of the exchange program between MPG and CAS for
a number of bilateral visits. 

\end{acknowledgements}



\bibliographystyle{aa}

\end{document}